\documentclass[11pt]{article}
\usepackage{graphicx}
\usepackage{epsfig}
\usepackage{amsfonts}
\usepackage{slashed}
\usepackage{mathrsfs}
\usepackage{amssymb}
\usepackage{amsthm,amsmath,amssymb}
\usepackage{hyperref}
\usepackage{graphicx}       
\usepackage{dcolumn}        
\usepackage{bm}             
\usepackage{amssymb}
\usepackage{amstext}
\usepackage{tensor}
\usepackage{cite}
\usepackage{subfig}
\usepackage{diagbox}
\usepackage[usenames]{color}
\usepackage{color}
\usepackage{graphics}
\usepackage{indentfirst}
\usepackage{booktabs}
\usepackage{float}
\usepackage{multirow}
\usepackage[section]{placeins}
\usepackage[capitalize]{cleveref}
\usepackage{array}
\newcolumntype{C}{>{8pt}c}

\usepackage{amssymb}

\date{\today}

\newcommand{\insertplot}[5]{\begin{figure}
		\hfill\hbox to 0.05in{\vbox to #5in{\vfill
				\inputplot{#1}{#4}{#5}}\hfill}
		\hfill\vspace{-.1in}
		\caption{#2}\label{#3}
\end{figure}}
\newcommand{\inputplot}[3]{
	\special{ps: plotfile #1}
}
\newcounter{fig}

\newcommand{\ee}{\end{equation}}
\newcommand{\eea}{\end{eqnarray}}
\newcommand{\be}{\begin{equation}}
\newcommand{\bea}{\begin{eqnarray}}

\begin{document}
	\title{\Large{\bf Rotating multistate Proca stars  }}
	\vspace{1.5truecm}

	\author{
		{\large }
        {\  Rong Zhang},
        {\  Long-Xing Huang},
		and
		{\ Yong-Qiang Wang\footnote{yqwang@lzu.edu.cn, corresponding author
		}}
		\\
		\\
		$^{1}${\small Lanzhou Center for Theoretical Physics, 
		}
		\\
		{\small
			Key Laboratory of Theoretical Physics of Gansu Province,}
		\\
		{\small
			School of Physical Science and Technology,  }
		\\
		{\small
			Lanzhou University,Lanzhou 730000, China}
		\\
		$^{2}${\small Institute of Theoretical Physics $\&$ Research Center of Gravitation, 
		}
		\\
		{\small
			Lanzhou University,Lanzhou 730000, China}
		\\
	}
	
	\date{August 2023}
	
	\maketitle

	\begin{abstract}
	In this paper, we construct the rotating multistate Proca stars (RMPSs) composed of two coexisting Proca fields including the ground state and first excited state. We consider the solution families of RMPSs in synchronized frequency and non-synchronized frequency scenarios, respectively. We discuss the characteristics of the matter fields changing with the synchronized frequency or non-synchronized frequency, and then we explore the relationship between the ADM mass $M$ and the frequency. We also analyze the characteristics of ADM mass $M$ as a function of the angular momentum $J$ simultaneously. Furthermore, we calculate the binding energy of RMPSs and analyze the stability of the solutions. Finally, we discuss the ergosphere of RMPSs. 

	\end{abstract}
	\newpage
	\section{INTRODUCTION}
    Boson stars are theoretical, self-gravitating solitons that can be thought of as macroscopic Bose-Einstein condensates of massive bosons. As a macroscopic astrophysical model, boson stars can play various potential roles in astrophysics. For example, they can be regarded as a possible components for dark matter \cite{Sharma:2008sc,Eby:2015hsq,Chen:2020cef,Gorghetto:2022sue}, the black hole mimickers\cite{Vincent:2015xta,Cunha:2016bjh}, or the potential sources of gravitational-waves \cite{Palenzuela:2017kcg,Bezares:2017mzk,Bezares:2018qwa,Jaramillo:2022zwg}.

    The concept of boson stars can be traced back to the 1950s when John Wheeler coupled electromagnetic fields with gravitational fields to construct stable particle-like solutions\cite{Wheeler:1955zz,Power:1957zz}. Wheeler named these stable particle-like solutions geons, but the solutions he obtained were ultimately found to be unstable. Despite attempts to obtain geons were unsuccessful, Kaup\cite{Kaup:1968zz} and Ruffini\cite{Ruffini:1969qy} achieved success by obtaining spherically symmetric, stable self-gravitating solitons through the use of massive scalar fields, replacing classical electromagnetic fields. These particle-like solutions are the earliest instances of boson stars. In the following decades, the researches on scalar boson stars were expanded to various scenarios, such as rotating boson stars \cite{Schunck:1996he,Yoshida:1997qf,Kleihaus:2005me,Li:2019mlk,Siemonsen:2020hcg}, cases in the anti-de Sitter spacetime\cite{Astefanesei:2003qy,Prikas:2004yw,Buchel:2013uba}, charged boson star\cite{Jetzer:1989av,Jetzer:1992tog,Brihaye:2014gua}, cases with self-interacting potentials\cite{Schunck:1999zu,Kling:2017hjm,Sanchis-Gual:2021phr}, and hairy black holes\cite{Herdeiro:2014goa,Herdeiro:2015waa,Herdeiro:2015gia,Herdeiro:2015tia,Cunha:2016bpi,Delgado:2020hwr}.

    On the other hand, researchs on vector boson stars started later. During the 1990s, N. Rosen\cite{Rosen:1994rq} coupled a massive vector field with a gravitational field, deriving the motion equation in spherically symmetric spacetime, but did not solve for stable particle-like solutions. It was not until 2015 that Brito et al. \cite{Brito:2015pxa} successfully solved for spherically symmetric vector boson stars and subsequently extending their study to the case with rotation. Such boson stars composed of vector bosons are also known as Proca stars. Proca stars share similarities with scalar boson stars and can also be generalized to various scenarios\cite{SalazarLandea:2016bys,Duarte:2016lig,Herdeiro:2017phl,Minamitsuji:2018kof,Delgado:2020hwr,Herdeiro:2023lze}. However,the stability properties of Proca stars diverge from those of scalar boson stars.\cite{Sanchis-Gual:2017bhw,Sanchis-Gual:2019ljs,DiGiovanni:2020ror,Herdeiro:2021lwl}. In particular, the ground state rotating Proca stars can remain stable under non-axisymmetric perturbations. Inspired by this property, Juan et al. \cite{CalderonBustillo:2020fyi} to study the collision of rotating Proca stars, comparing the simulated gravitational wave data with observations from GW190521. The results showed that the gravitational wave waveform resulting from Proca star collisions exhibited better consistency with observational data than quasi-circular binary black hole merger models.

    Some recent studies\cite{Ma:2023vfa,Pombo:2023sih} reveal that multifield solutions resulting from the coupling of Proca fields and scalar fields exhibit distinct properties compared to single-field Proca stars. These studies, along with related works\cite{Bernal:2009zy,Li:2020ffy,Guo:2020tla,Sanchis-Gual:2021edp,Sun:2022duv,Zeng:2023hvq,Liang:2022mjo,Huang:2023glq} indicate that the physical properties of the single-field boson stars, such as mass, angular momentum, binding energy, and dynamical stability, differ from those of the corresponding multifield boson stars. Additionally, multifield boson stars possess unique properties absent in their single-field counterparts. For example, some studies \cite{Liang:2022mjo,Ma:2023vfa,Huang:2023glq} found that the number of branches of the multifield solutions are affected by frequency and mass of matter fields. Presently, multifield model studies of Proca fields are limited to spherically symmetric conditions, yet rotation is a ubiquitous phenomenon in nature. Hence, the primary aim of this paper is to extend the research on Proca stars to the rotating multifield solutions.

    The organization of this paper is as follows: In Section 2, we introduce the rotating multistate Proca star model (RMPSs), including the motion equations and the ansatz for the metric and Proca field. We also present the formulas for several important quantities. Section 3 covers the boundary conditions and the numerical methods used for calculations. Section 4 presents the numerical results of RMPSs, considering both synchronized frequency $(\omega_{0}=\omega_{1})$ and non-synchronized frequency $(\omega_{0}\neq\omega_{1})$. We compute several important quantities and analyze the ergosphere of multifield solutions. Finally, in Section 5, we summarize our research findings and discuss potential directions for future work.
	
	\section{THE MODEL SETUP}
	\label{sec2}
	We consider a model of two Proca fields with minimal coupling to Einstein gravity in 3+1 dimensional spacetime. The action is:
    \begin{equation}
       S=\int{d}^4x\sqrt{-g}\left( \frac{R}{16\pi G}+\mathcal{L}^{0}+\mathcal{L}^{1}\right), \label{Action} 
    \end{equation} 
    where $G$ is the gravitational constant and $R$ is the Ricci scalar. These two Proca fields are represented by the symbol $i$ $(i=0,1)$, hence $\mathcal{L}^{i}$ denotes the Lagrangian density of them. Which reads:
   \begin{equation}
       \mathcal{L}^{i}=-\frac{1}{2}\mathcal{F}^{(i)}_{\alpha \beta}\bar{\mathcal{F}}^{(i)}_{\alpha \beta}-\frac{1}{2}\mu_{i}^{2}\mathcal{A}^{(i)}_{\alpha} \bar{\mathcal{A}}^{(i)\,\alpha}.\label{Lagrangian}
    \end{equation}
   Here $\mathcal{A}$ is the potential of the Proca field, $\mathcal{F}$ represents the field strength, and their relationship is $\mathcal{F}=d\mathcal{A}$. Their complex conjugates are denoted by $\bar{\mathcal{A}}$ and $\bar{\mathcal{F}}$.\par

   By varying this action Eq. (\ref{Action}), we can obtain the Einstein field equation and the Proca field equations:
   \begin{align}      
      R_{{\alpha\beta }}-\frac{1}{2}R{g}_{{\alpha\beta }}&=8\pi G (T^{0}_{\alpha \beta}+T^{1}_{\alpha \beta}), \label{eq-Einstein}\\
     \nabla_\alpha\mathcal{F}^{(i)\,\alpha\beta}&=\mu_{i}^2\mathcal{A}^{\beta\,(i)}, \label{eq-Proca}
   \end{align} 
where $\mu_i$ is the mass of Proca fields, and the energy-momentum tensor reads:
   \begin{equation}
       T^{i}_{\alpha \beta}=-\mathcal{F}^{(i)}_{\sigma (\alpha}\bar{\mathcal{F}}_{\beta {)}}^{\sigma\,(i)}-\frac{1}{4}g_{\alpha \beta}\mathcal{F}^{(i)}_{\sigma \tau}\bar{\mathcal{F}}^{\sigma \tau\,(i)}+\mu_i^2\left[ \mathcal{A}^{(i)}_{{(}\alpha}\bar{\mathcal{A}}^{(i)}_{\beta {)}}-\frac{1}{2}g_{\alpha \beta}\mathcal{A}^{(i)}_{\sigma}\bar{\mathcal{A}}^{\sigma\,{(i)}} \right].
    \end{equation}

   In our study, we adopt the spheroidal coordinates coordinates $\left(t,r,\theta,\varphi\right)$, for which the metric will be more convenient to calculate
   \begin{equation}
        ds^2=- e^{2F_0(r,\theta)}dt^2+e^{2F_1(r,\theta)}(dr^2+r^2 d\theta^2) +e^{2F_2(r,\theta)}r^2 \sin^2\theta \left(d\varphi-\frac{W(r,\theta)}{r}dt\right)^2. \label{ansatz-metric}
    \end{equation} \par
   In the spacetime described by metric (\ref{ansatz-metric}), we use the following ansatz for the Proca field:
  \begin{equation}
        \mathcal{A}_{(n)}^{(i)}=\left(
\frac{H_{1(n)}^{i}}{r}dr+H_{2(n)}^{i}d\theta+i H_{3(n)}^{i} \sin \theta d\varphi + iV_{(n)}^{i}dt   
\right) e^{i(m_{i} \varphi-\omega_{i} t)},\,\ m_{i}=\pm{1},\pm{2},\dots, \label{ansatz-mass} 
   \end{equation}
where the subscript $(n)$ represents the number of nodes of each matter fields, which is the sum of the radial node number $n_{r}$ and the angular node number $n_\theta \,\ (n=n_{r}+n_{\theta})$. Inspired by hydrogen orbitals, we call the Proca field with $n=0$ the ground state, where all matter field functions have no nodes, and we call the Proca field with $n>0$ the excited states. The functions $(H_{j}^{i},V^{i}) \,\ (j=1,2,3)$ are real functions depending on $(r,\theta)$, and $m_{i}$ and $\omega_{i}\,\ (i=0,1)$ respectively denotes the azimuthal harmonic index and the frequency of the Proca field. When $\omega_{0}$ and $\omega_1$ synchronously change $(\omega_{0}=\omega_{1}=\omega)$, the frequency $\omega$ of these Proca fields is called the synchronized frequency, while $\omega_{1}$ is called the non-synchronized frequency when $\omega_{0}$ does not change synchronously with $\omega_1$ $(\omega_{0}\neq\omega_{1})$. 
 
    In our research, we are interested in several important quantities of RMPSs, including their ADM mass, angular momentum, and binding energy. Firstly, the ADM mass of RMPSs can be obtained by integrating the Komar energy density over any arbitrary spacelike slice $\Sigma$:
   \begin{equation}
    M=\int_\varSigma{drd\theta d\varphi \sqrt{-g}\left( 2 T_{t}^{t}-T_{\alpha}^{\alpha} \right),} \label{eq-KormarEnergy}
   \end{equation}
    where the integrated function $\left( 2T_{t}^{t}-T_{\alpha}^{\alpha} \right)$ is the Komar energy density.

   Secondly, similar to Eq. (\ref{eq-KormarEnergy}), the angular momentum of RMPSs can be obtained by integrating Komar angular momentum density over $\Sigma$
    \begin{equation}
        J=\int_\varSigma{drd\theta d\varphi \sqrt{-g} T_{\varphi}^{t}}. \label{eq-AngularMomentum}
    \end{equation}
   Since the action of the matter fields possesses a global $U(1)$ symmetry under the transformation ${A}_{\alpha}\rightarrow e^{i\gamma}{A}_{\alpha}$, with $\gamma$ constant. According to Noether's theorem, this implies the existence of conserved 4-currents, which reads:
    \begin{equation}
        j^{\alpha (i)}=\frac{1}{2}\left[\bar{\mathcal{F}}^{(i)\,\alpha \beta}\mathcal{A}^{(i)}_{\beta}-\mathcal{F}^{(i)\,\alpha\beta}{\mathcal{\bar{A}}}^{(i)}_{\beta}\right].\label{current}
    \end{equation}
    According to the calculations in Ref.\cite{Herdeiro:2016tmi}, the Komar angular momentum density can be expressed as follows:
        \begin{equation}
        \begin{split}
         T_{\varphi}^{t}=m_{0} j^{t\,(0)}+m_{1} j^{t\,(1)}+\nabla_{\alpha}\mathcal{P}^{\alpha}, \label{eq-Ttvarphi}
        \end{split}
    \end{equation}
where $\mathcal{P}^{\alpha}=\mathcal{A}^{(0)}_{\varphi}\Bar{\mathcal{F}}^{{(0)}\,\alpha t}+{\mathcal{\bar{A}}^{(0)}_{\varphi}}\mathcal{F}^{{(0)}\,\alpha t}+\mathcal{A}^{(1)}_{\varphi}\Bar{\mathcal{F}}^{{(1)}\,\alpha t}+{\mathcal{\bar{A}}^{(1)}_{\varphi}}\mathcal{F}^{{(1)}\,\alpha t}$, and $j^{t(i)}$ is the time component of the conserved current.

    From the Proca equation [\ref{eq-Proca}], we know that $\nabla_\alpha j_{(i)}^\alpha=0$. Therefore, if we integrate the time component of the conserved current over any spacelike slice $\Sigma$, we can obtain the conserved Noether charge:
    \begin{equation}
        Q^{i}=\int_\Sigma d^3x \sqrt{-g} j^{t (i)}.
    \end{equation}   
And the angular momentum of RMPSs can be obtained by the following equation from Eq. (\ref{eq-AngularMomentum}) and Eq. (\ref{eq-Ttvarphi}):
   \begin{equation}
       J=m_{0} Q^{0}+m_{1} Q^{1}.
   \end{equation}\par

    On the other hand, the mass $M$ and the total angular momentum $J$ can also be obtained from the asymptotic behavior of the metric functions at infinity:
   \begin{equation}
    \begin{split}
        g_{tt}|_{r=\infty}&=-e^{2F_0}+e^{2F_2}W^2\sin^2{\theta}=-1+\frac{2GM}{r},\\
        g_{\varphi t}|_{r=\infty}&=-e^{2F_0}Wr\sin^2{\theta}=-\frac{2GJ}{r}\sin^2\theta.
    \end{split}
   \end{equation}

    In addition, we are interested in the stability of the system. It is generally accepted that the system is stable when the binding energy is negative. Therefore, in this article, we will calculate the binding energy $E_m$ of the RMPSs, which can be obtained from the following equation:
    \begin{equation}
       E_m=M-\mu_{0}Q^0-\mu_{1}Q^1.
    \end{equation}\par 
\section{BOUNDARY CONDITIONS AND NUMERICAL IMPLEMENTATION}
\label{sec3}
    To solve the partial differential equations based on the ansatz (\ref{ansatz-metric}) and (\ref{ansatz-mass}), it is necessary to give the boundary conditions of the unknown functions. Considering the properties of RMPSs, we will set the boundary conditions according to the steps given in reference Ref.\cite{Brito:2015pxa}. Since the RMPSs solution is asymptotically flat, the field functions should satisfy the following boundary conditions at infinity:
    \begin{equation}
    F_{l}|_{r=\infty}=W|_{r=\infty}=H_j^{i}|_{r=\infty}=V^{i}|_{r=\infty}=0,
    \end{equation}
    where $l=0,1,2$, $j=1,2,3$. And at $r=0$, they satisfy:
    \begin{equation}
        \partial_{r}F_l|_{r=0}=W|_{r=0}=H_j^{i}|_{r=0}=V^{i}|_{r=0}=0.
    \end{equation}

    For the case of $m_{i}=1$, the following boundary conditions are compatible with the solutions at $\theta=0,\pi$:
    \begin{equation}
\partial_{\theta}F_l|_{\theta=0,\pi}=\partial_{\theta}W|_{\theta=0,\pi}= H_1^{i}|_{\theta=0,\pi}=\partial_{\theta}H_2^{i}|_{\theta=0,\pi}=\partial_{\theta}H_3^{i}|_{\theta=0,\pi}=V^{i}|_{\theta=0,\pi}=0.
    \end{equation}

    By introducing the following equatorial boundary conditions, all solutions in this paper are symmetric with respect to a reflection along the equatorial plane:
    \begin{equation}
        \partial_{\theta}H_1^{i}|_{\theta=\pi/2}=H_2^{i}|_{\theta=\pi/2}=\partial_{\theta}H_3^{i}|_{\theta=\pi/2}=\partial_{\theta}V^{i}|_{\theta=\pi/2}=0. \label{eq-boundary}
    \end{equation}

    In addition, for the case of $m_{i}\neq1$, the boundary conditions of $H_2^{i}$ and $H_3^{i}$ will be different from Eq. (\ref{eq-boundary}), while the azimuthal harmonic index of all RMPSs in this paper satisfies $m_{i}=1$.
    For convenience of calculation, we can introduce the following dimensionless quantities:
    \begin{equation}
        \begin{split}
       \widetilde{M}=\frac{M\mu_0}{M_{PL}^2},\quad\widetilde{J}=\frac{J \mu_0}{M_{PL}^2}&,\quad\widetilde{E}_m=\frac{{E_m}{\mu_0}}{M_{PL}^2},\\
       \widetilde{r}=r\mu_0,\quad \widetilde{{H}}_j^{i}=\frac{\sqrt{2 \pi }}{M_{PL}} H_j^{i},\quad\widetilde{V}^{i}=&\frac{\sqrt{2 \pi }}{M_{PL}} V^{i},\quad \widetilde{\omega}_{i}=\frac{\omega_{i}}{\mu_0}, \quad \widetilde{\mu}_{i}=\frac{\mu_{i}}{\mu_0},
    \end{split}
    \end{equation}
where $M_{PL}\equiv 1/\sqrt{G}$ is the Planck mass. Furthermore, we introduce the following conformal transformation for convenience:
    \begin{equation}
        x = \frac{\widetilde{r}}{1+\widetilde{r}} \label{eq-radius}.
    \end{equation}
    It can be easily obtained that transformation (\ref{eq-radius}) transforms the range of radial coordinates $\widetilde{r}\in[0,+\infty )$ into $x\in[0,1)$. Finally, for convenience, these dimensionless physical quantities will still be expressed using original symbols in subsequent sections.

	\section{NUMERICAL RESULTS}
		\label{sec4}		
    In this section, we present results of the RMPSs, along with an analysis of their several key properties. Based on the relationship between the frequencies of the two Proca fields, we categorize the multistate solutions into synchronized frequency $(\omega_{0}=\omega_{1}=\omega)$ and non-synchronized frequency $(\omega_{0}\neq\omega_{1})$ RMPSs. They are composed of the ground state Proca field $(n=0)$ and the first excited state Proca field $(n=1)$, both with azimuthal harmonic indexes equal 1 $(m_{0}=m_{1}=1)$. In our study, each matter field function in the excited state has a radial node $(n=n_{r}=1)$ and no node in the angular direction $(n_{\theta}=0)$. As no solutions with a lower total number of nodes have been found at present, we refer to this type of solution as the first excited state. In addition, to compare the properties of RMPSs with those of single-field Proca stars, we also calculated the ground state Proca star and the first excited state Proca star. We denote the ground state Proca star as $P_0$, the first excited state Proca star as $P_1$, and RMPSs as $P_0P_1$.

    Our solving process is based on the finite element analysis methods, which solves weak form of partial differential equations to obtain numerical results. The iteration process follows the Newton-Raphson method. In the integration area $x\in[0,1]$ and $\theta\in[0,\pi/2]$, the grid is taken as $N_r\times N_\theta$. In general, we set $N_{r}=160$ and $N_\theta=120$. Using the above numerical method, the relative error of the obtained results is generally below $1\times 10^{-4}$.
    \subsection{Synchronized frequency}

   \begin{figure}[b!]
        \centering
        \subfloat{
        \includegraphics[width=7cm]{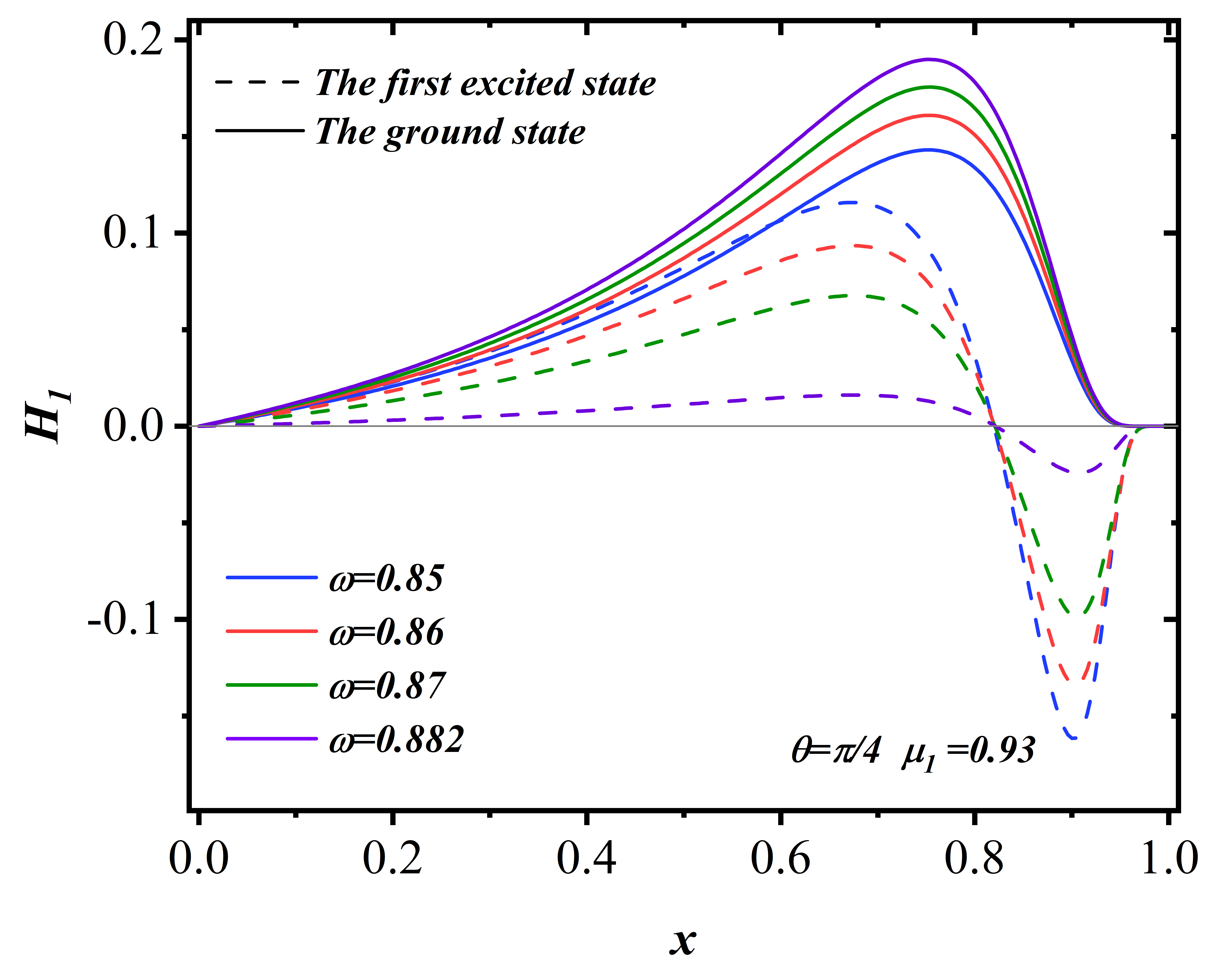}
        \label{fig:syn-field01}}
        \subfloat{
        \includegraphics[width=7cm]{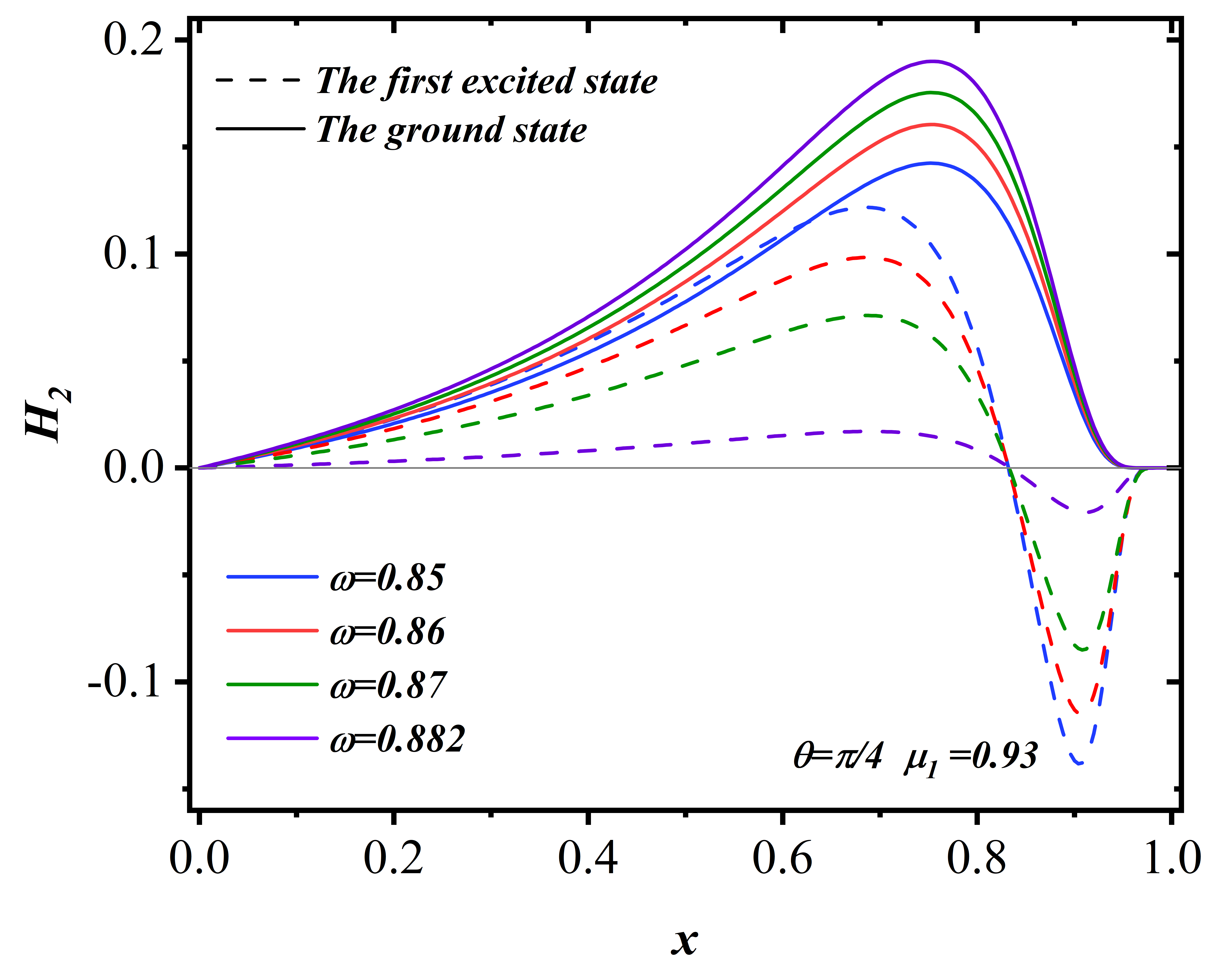}
        \label{fig:syn-field02}}
        \quad
        \subfloat{
        \includegraphics[width=7cm]{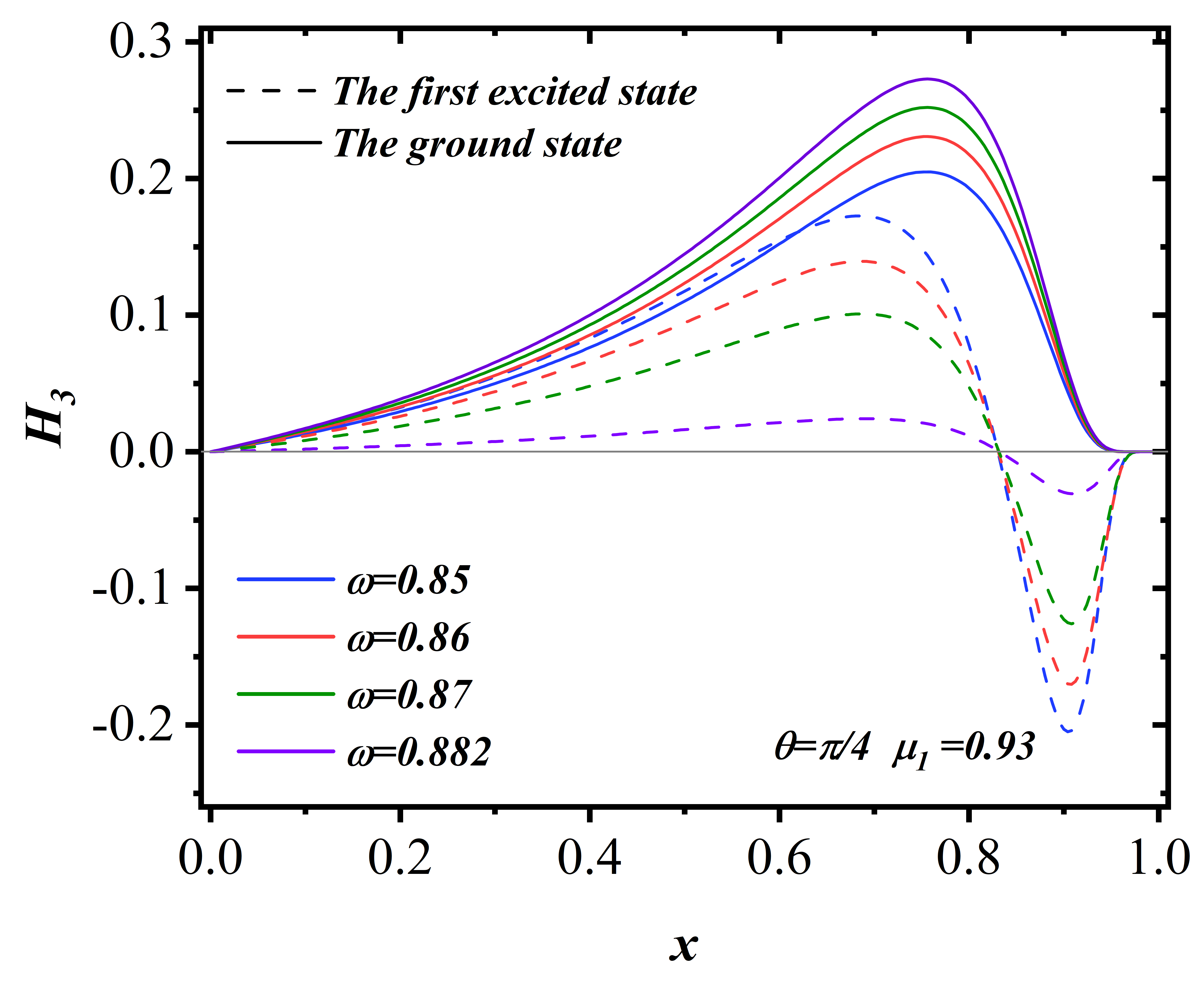}
        \label{fig:syn-field03}}
        \subfloat{
        \includegraphics[width=7cm]{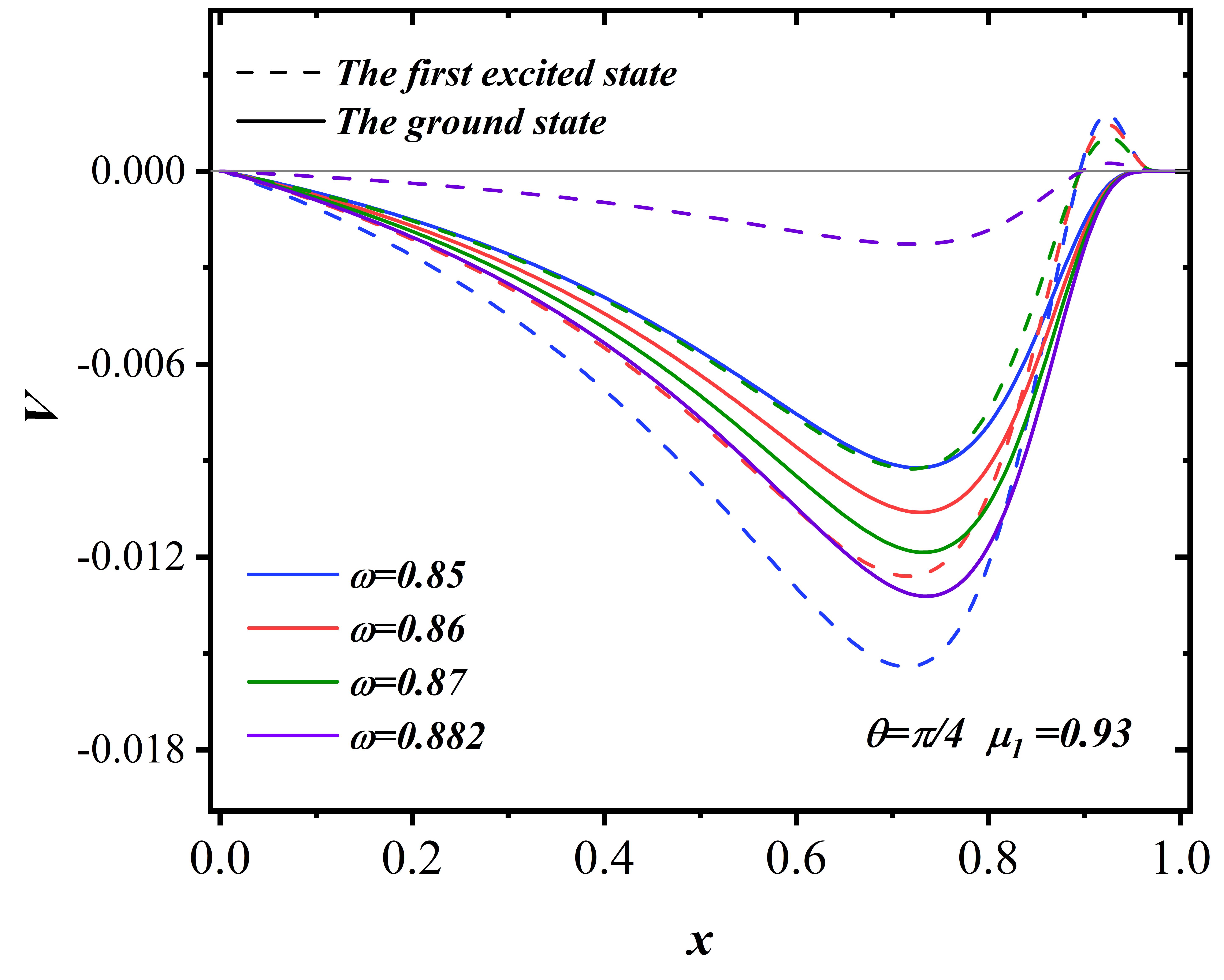}
        \label{fig:syn-field04}}
        \caption{The matter fields distribution of the ground state and first excited state with respect to $x$ for severals different value of the synchronized frequency $\omega$ when $\theta=\frac{\pi}{4}$. The solid and dashed lines represent the ground state and first excited state, respectively, and different colors represent $\omega=0.85$ (blue), $\omega=0.86$ (red), $\omega=0.87$ (green), and $\omega=0.882$ (purple), respectively. }
        \label{fig:syn-field}
    \end{figure} 
    This subsection discusses the RMPSs when the ground state frequency and the excited state frequency $\omega_1$ synchronously change. For this case, the masses of the two Proca fields must be different $(\mu_0 \neq \mu_1)$, otherwise the multistate solution will not be obtained. In this subsection, we calculate the synchronized frequency RMPSs under a fixed value of $\mu_1$ ($\mu_1\neq 1$). 

    \begin{figure}[!b]
        \centering
        \subfloat{
        \includegraphics[width=7cm]{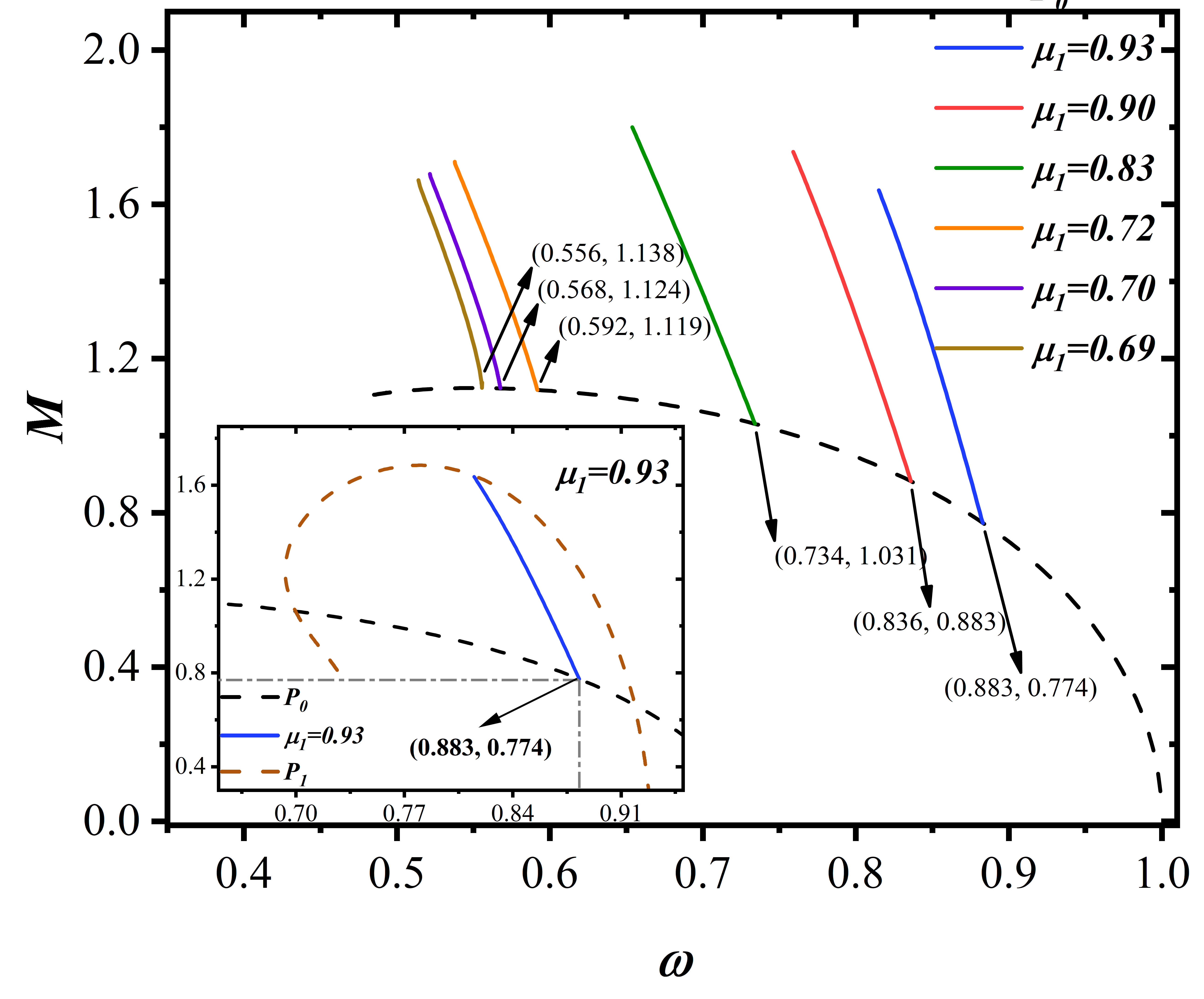}
        \label{fig:syn-J-M01}}
        \centering
        \subfloat{
        \includegraphics[width=7cm]{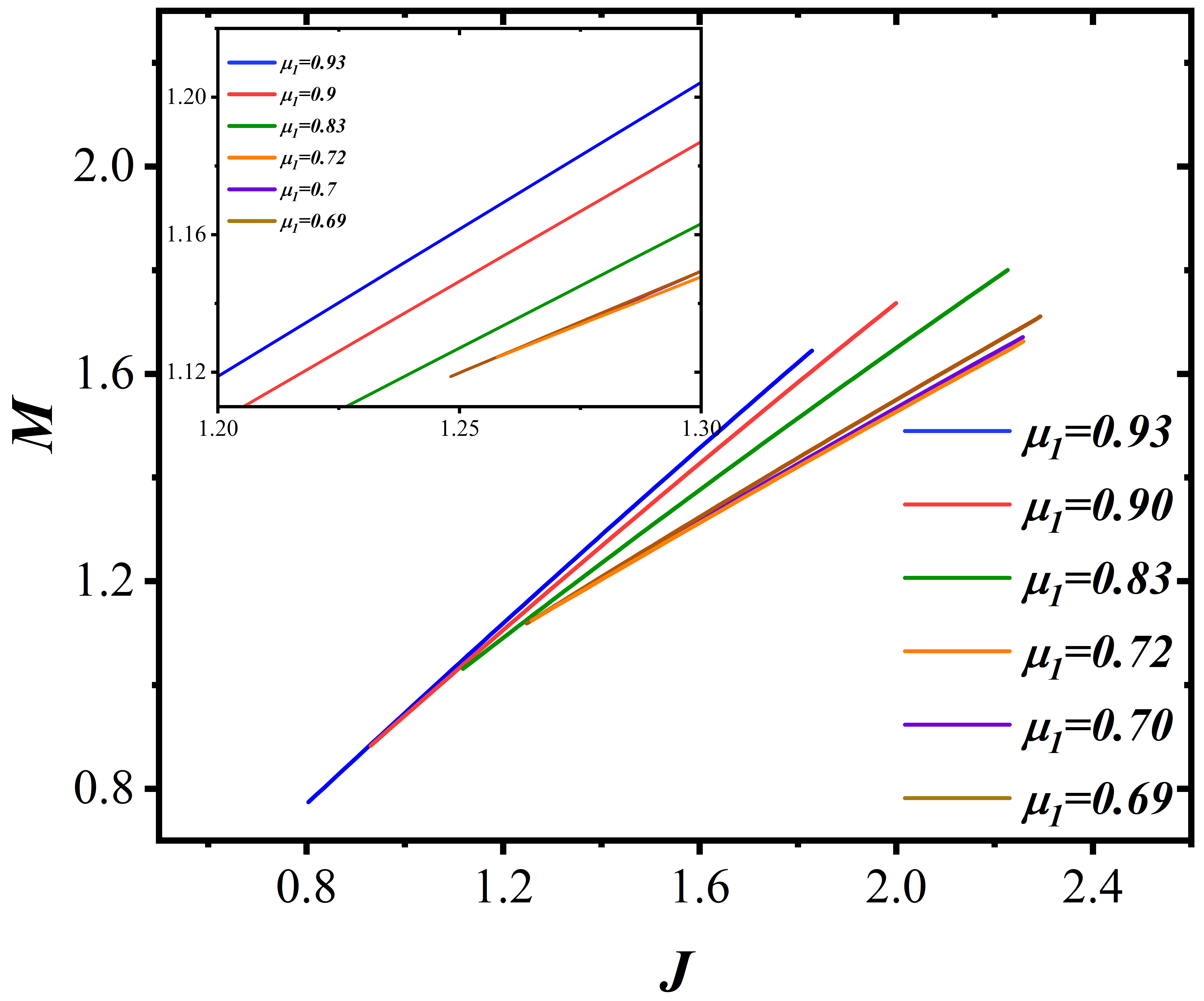}
        \label{fig:syn-J-M02}}
        \caption{Left panel: The ADM mass $M$ of synchronized RMPSs (solid lines) as a fuction of synchronized frequency $\omega$ for several values of $\mu_{1}$. The different clorful solid lines represent $\mu_{1}=0.69$ (brown), 0.70 (purple), 0.72 (orange), 0.83 (green), 0.90 (red) and 0.93 (blue), respectively. While the dashed lines represent the ground state Proca stars (black) and the first excited state Proca stars (brown), respectively. Right panel: The ADM mass $M$ of synchronized RMPSs as a function of angular momentum $J$ for $\mu_{1}=0.69, 0.70, 0.72, 0.83, 0.90, 0.93$. }
        \label{fig:syn-J-M}
    \end{figure}

    Fig. \ref{fig:syn-field} depicts the distribution of the Proca field with respect to $x$ at $\theta=\pi/4$, where the mass of the first excited state Proca field is set to 0.93 $(\mu_{1}=0.93)$. In the figure, different colors correspond to $\omega=0.85$ (blue), $\omega=0.86$ (red), $\omega=0.87$ (green), and $\omega=0.882$ (purple), respectively. And the solid lines represent the ground state, the dashed lines represent the first excited state. From this figure, it can be seen that in the radial direction, the excited state field functions has an intersection, while the ground state field functions does not intersect. Moreover, as $\omega$ increases, the values of $|{H}_j^{0}|_{max}$ and $|{V}^{0}|_{max}$ (the maximum absolute values of the four field functions of the ground state) increase, while those of the first excited state field functions $|{H}_j^{1}|_{max}$ and $|{V}^{1}|_{max}$ decrease. When the synchronized frequency increases to its maximum value, the first excited state Proca field in RMPSs vanishes, and multifield solutions degenerate into the ground state Proca star $P_0$. Conversely, when the synchronized frequency decreases to its minimum value, RMPSs degenerate into the first excited state Proca star $P_1$.
    \begin{table}[!t] 
	\centering 
	\begin{tabular}{|c||c|c|c|}
		\hline
		 &$\omega$ &$M$ &$J$     \\
		\hline
		$\mu_{1}=0.69$ & $0.514\sim0.556$ & $1.138\sim1.663$ & $1.258\sim2.258$\\ \hline
		$\mu_{1}=0.70$ & $0.522\sim0.568$ & $1.126\sim1.671$ & $1.261\sim2.257$\\ \hline
		$\mu_{1}=0.72$ & $0.538\sim0.592$ & $1.119\sim1.711$ & $1.248\sim2.293$\\ \hline
		$\mu_{1}=0.83$ & $0.654\sim0.734$ & $1.031\sim1.800$ & $1.118\sim2.226$\\ \hline
        $\mu_{1}=0.90$ & $0.758\sim0.836$ & $0.883\sim1.736$ & $0.930\sim1.999$\\ \hline
        $\mu_{1}=0.93$ & $0.814\sim0.883$ & $0.774\sim1.645$ & $0.804\sim1.828$\\
		\hline
	\end{tabular}
 	\caption{The range of synchronized frequency $\omega$, ADM mass $M$, and angular momentum $J$. }
	\label{tab:syn-M-J}
    \end{table}

    To further discuss the properties of synchronized frequency RMPSs, we analyze the ADM mass $M$ and angular momentum $J$ of the synchronized frequency RMPSs. Fig. \ref{fig:syn-J-M} depicts the ADM mass $M$ vs. synchronized frequency $\omega$ (left) and $M$ vs. $J$ (right) for six different values of $\mu_{1}$. To better illustrate the relationship between RMPSs and single-field Proca stars, the left panel of Fig. \ref{fig:syn-J-M} also depicts the ADM mass $M$ of the ground state Proca star $P_0$ ($\mu=1$) (black dashed line) and the first excited state Proca star $P_1$ ($\mu=0.93$) (brown dashed line) as a function of their frequencies.

    From the left panel of Fig. \ref{fig:syn-J-M}, it can be seen that under a fixed value of $\mu_{1}$, RMPSs exist in a certain range of the synchronized frequency $\omega$, which is similar to the case of single-field Proca stars. The difference is that the graph of the first excited state Proca star has a behavior of spiraling to the center, which is nonexistent in the case of RMPSs. We can also observe that as $\omega$ increases, M is to decrease monotonically. When $\omega_1$ increases to its maximum value, RMPSs degenerate into the $P_0$, with the graph of the $P_0P_1$ intersect with $P_0$ at $(0.556,1.138)$, $(0.568,1.126)$, $(0.592,1.119)$, $(0.734,1.031)$, $(0.836,0.883)$, and $(0.883,0.774)$, respectively. When $\omega$ of RMPSs tends to its minimum value, the curve intersects with $P_1$, as shown in the embedded figure. In addition, according to the right panel of Fig. \ref{fig:syn-J-M}, ADM mass of RMPSs increases monotonically with increasing angular momentum $J$ which is not the same as the result of single-field solutions in Ref.\cite{Herdeiro:2020jzx}.
    \begin{figure}[!b]
        \centering
        \includegraphics[width=7.5cm]{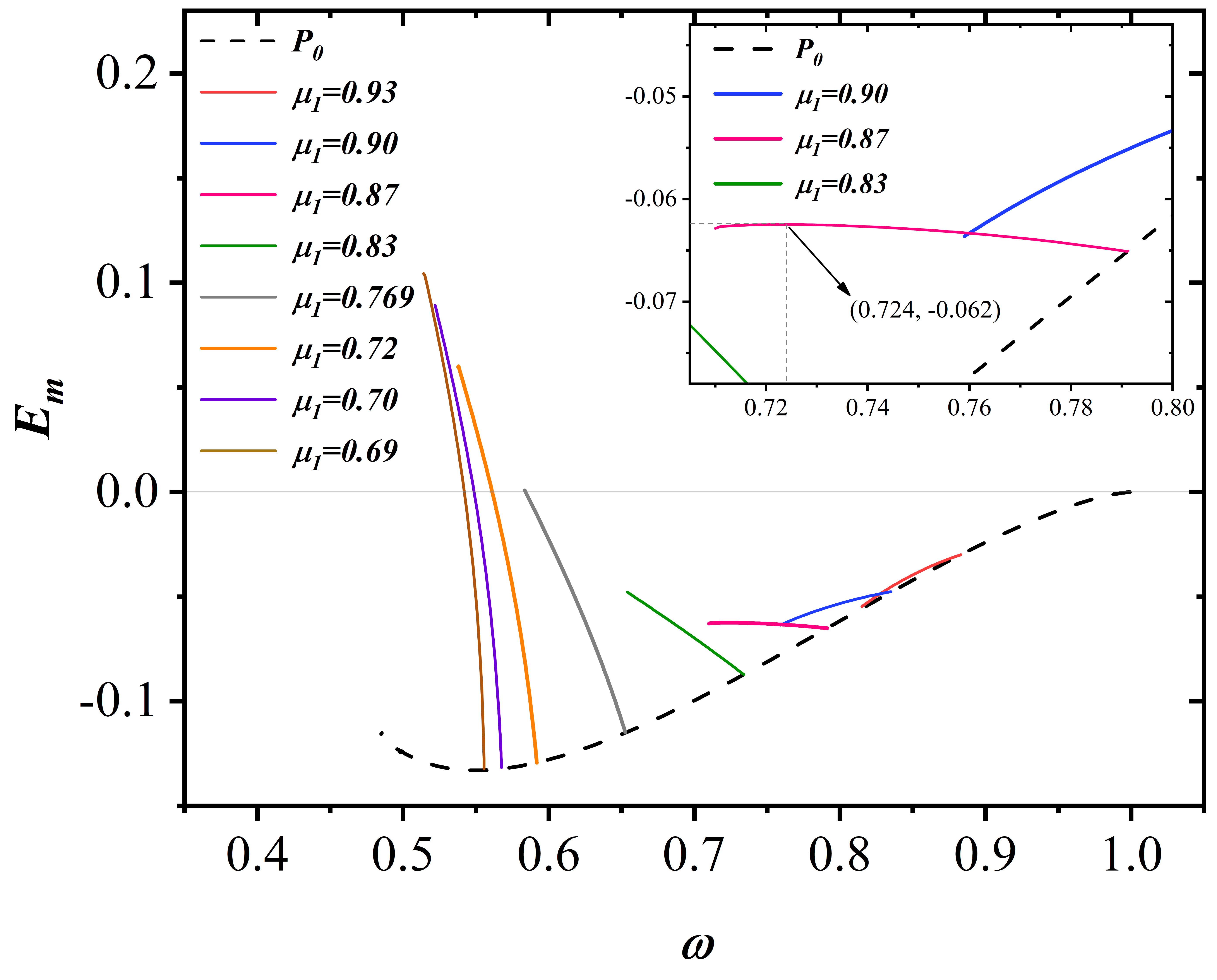}
        \caption{The binding energy $E_m$ of RMPSs with respect to synchronized frequency $\omega$. While the black dashed line represent binding energy $E_0$ of the ground state Proca stars. The different clorful solid lines represent $\mu_{1}=0.69$ (brown), 0.70 (purple), 0.72 (orange), 0.769 (grey), 0.83 (green), 0.87 (pink), 0.90 (blue) and 0.93 (red), respectively. All solutions have $\mu_{0}=1$ and $m_0=m_1=1$.}
        \label{fig:syn-E}
    \end{figure}

   Furthermore, to better analyze how the ADM mass and angular momentum of RMPSs change with $\mu_{1}$, we show the ranges of synchronized frequency, ADM mass, and angular momentum of RMPSs in Tab. \ref{tab:syn-M-J}. It can be seen that as $\mu_{1}$ increases, the minimum ADM mass of RMPSs are to decrease, while the maximum and minimum values of the synchronized frequency are to increase.

    Next, we investigate the binding energy of the synchronized frequency RMPSs. Fig. \ref{fig:syn-E} show the binding energy of RMPSs as a function of $\omega$ for several values of $\mu_1$. To facilitate comparison, we also plot the binding energy of the ground state Proca star as a fuction of its frequency. The different solid lines represent $\mu_{1}=0.69$ (brown), 0.70 (purple), 0.72 (orange), 0.769 (grey), 0.83 (green), 0.87 (pink), 0.90 (blue) and 0.93 (red), respectively, while the dashed line represents the ground state Proca star.

    \begin{table}[!t]
	\centering 
	\begin{tabular}{|c||c|c|c|c|c|}
		\hline
		 & $\omega$ & $E_m$ & $\omega^{min}$ & $\omega^{max}$ & $\omega^{0}$ \\
		\hline
		$\mu_{1}=0.69$ & $0.514\sim0.556$ & $-0.132\sim0.104$ &0.556 &0.514 &0.542\\ \hline
		$\mu_{1}=0.70$ & $0.522\sim0.568$ & $-0.132\sim0.090$ &0.568 &0.522 &0.549\\ \hline
		$\mu_{1}=0.72$ & $0.538\sim0.592$ & $-0.129\sim0.060$ &0.592 &0.538 &0.562\\ \hline
		$\mu_{1}=0.769$ &$0.584\sim0.653$ &  $-0.115\sim0.001$ &0.584 &0.653 &0.584\\ \hline
		$\mu_{1}=0.83$ & $0.654\sim0.734$ & $-0.087\sim-0.048$ &0.654 &0.734 &---\\ \hline
		$\mu_{1}=0.87$ & $0.710\sim0.791$ & $-0.065\sim-0.063$ &0.724 &0.791 &---\\ \hline
        $\mu_{1}=0.90$ & $0.758\sim0.836$ & $-0.064\sim-0.048$ &0.758 &0.836 &---\\ \hline
        $\mu_{1}=0.93$ & $0.814\sim0.883$ &$-0.055\sim-0.030$ &0.814 &0.883 &---\\
		\hline
	\end{tabular}
    \caption{The range of the synchronized frequency and binding energy, and the values of the synchronized frequency corresponding to the maximum value, minimum value of $E_m$, and $E_m=0$ ($\omega^{max}$, $\omega^{min}$ and $\omega^{0}$).  }
    \label{tab:syn-E}
    \end{table}
    \begin{figure}[b!]
        \centering
        \subfloat{\includegraphics[width=7cm]{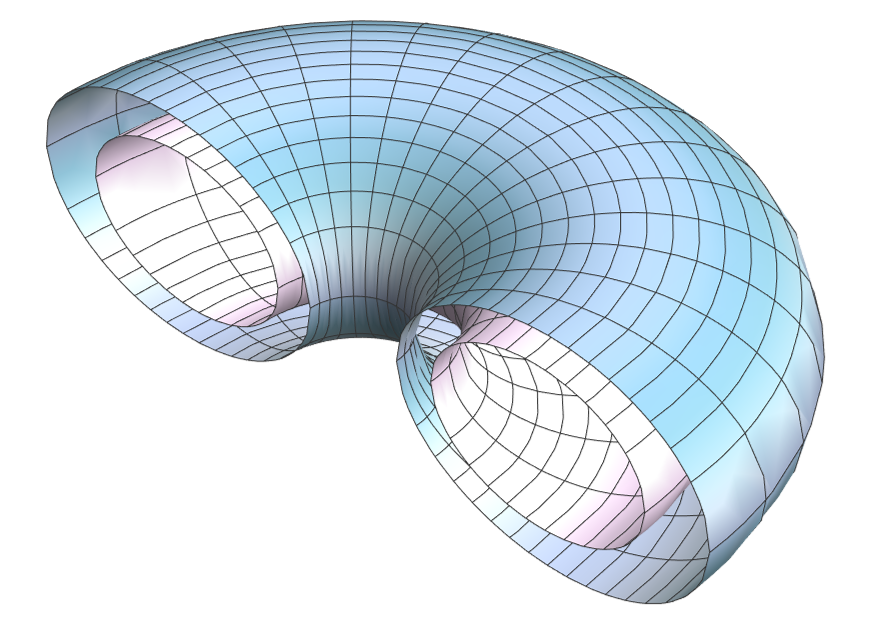}}
        \subfloat{\includegraphics[width=7cm]{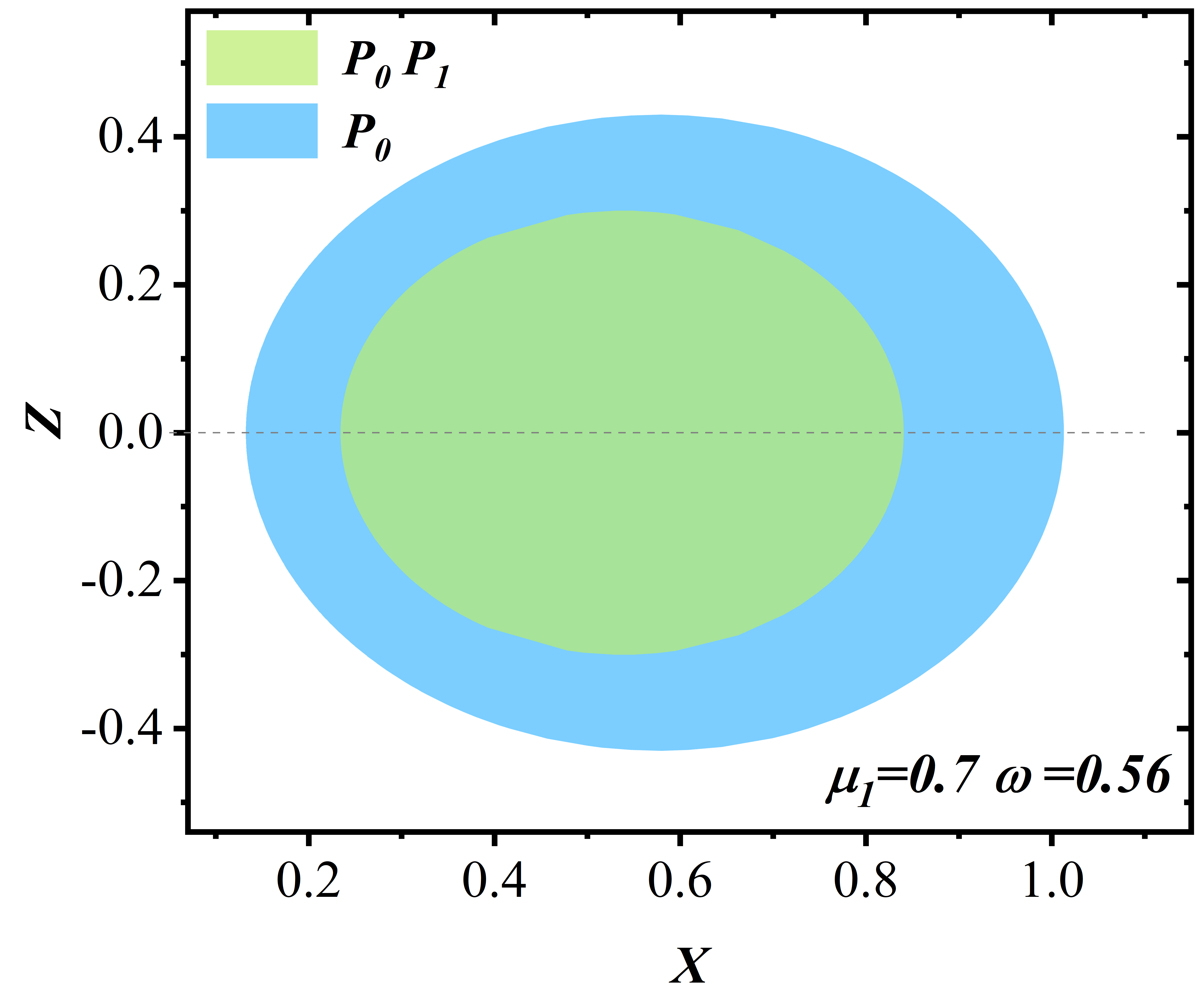}}
        \caption{Left: The ergosurface of ground state Proca stars (blue) and RMPSs (pink) in Cartesian coordinates. Right: The cross-sectional view of the ergosphere of the ground state Proca stars (blue) and RMPSs (green) in the $XZ$ plane. }
        \label{fig:Ergosphere-0.56}
    \end{figure}
        \begin{figure}[htbp]
		\centering
        \subfloat{
			\includegraphics[width=6.8cm]{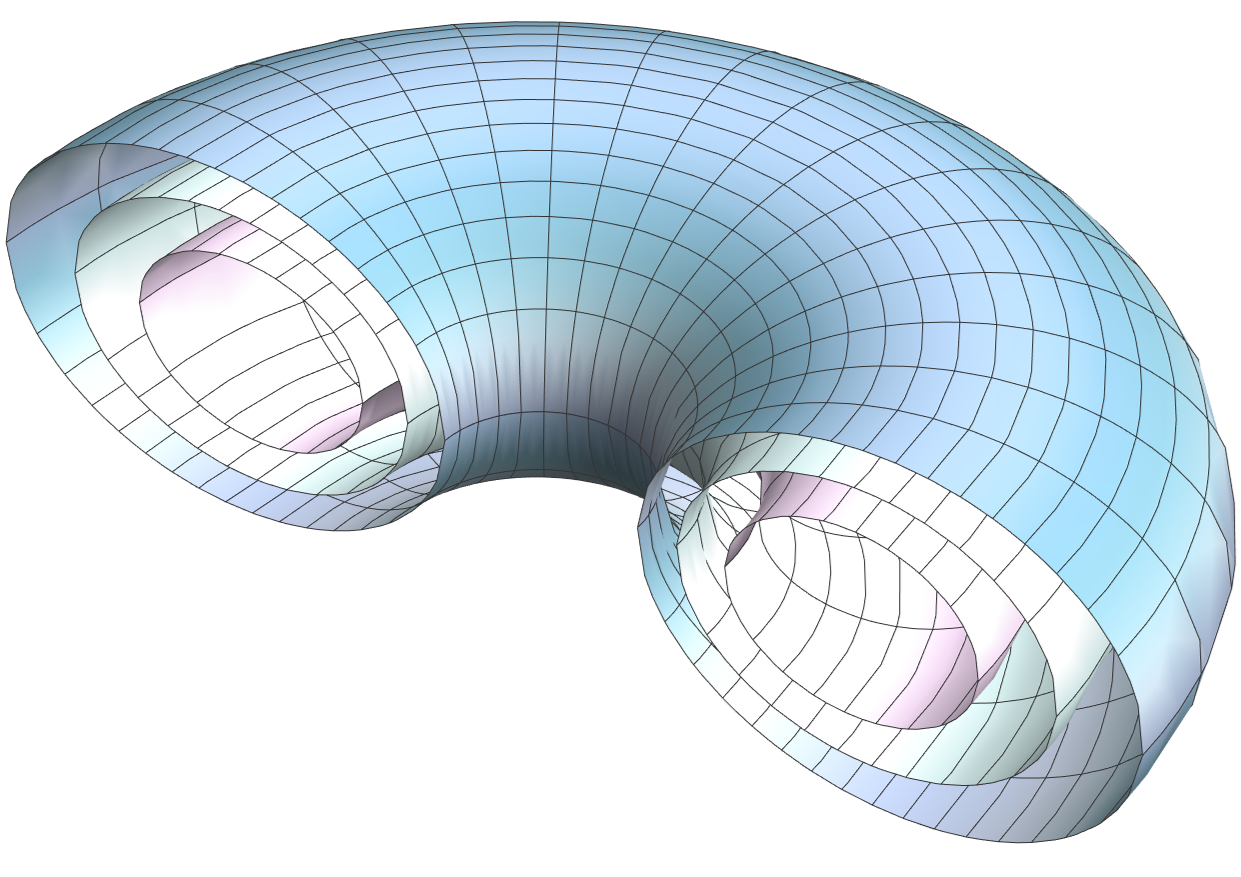}
   \label{fig:syn-Ergosphere01}} 	
      \quad
		\subfloat{
			\includegraphics[width=6.8cm]{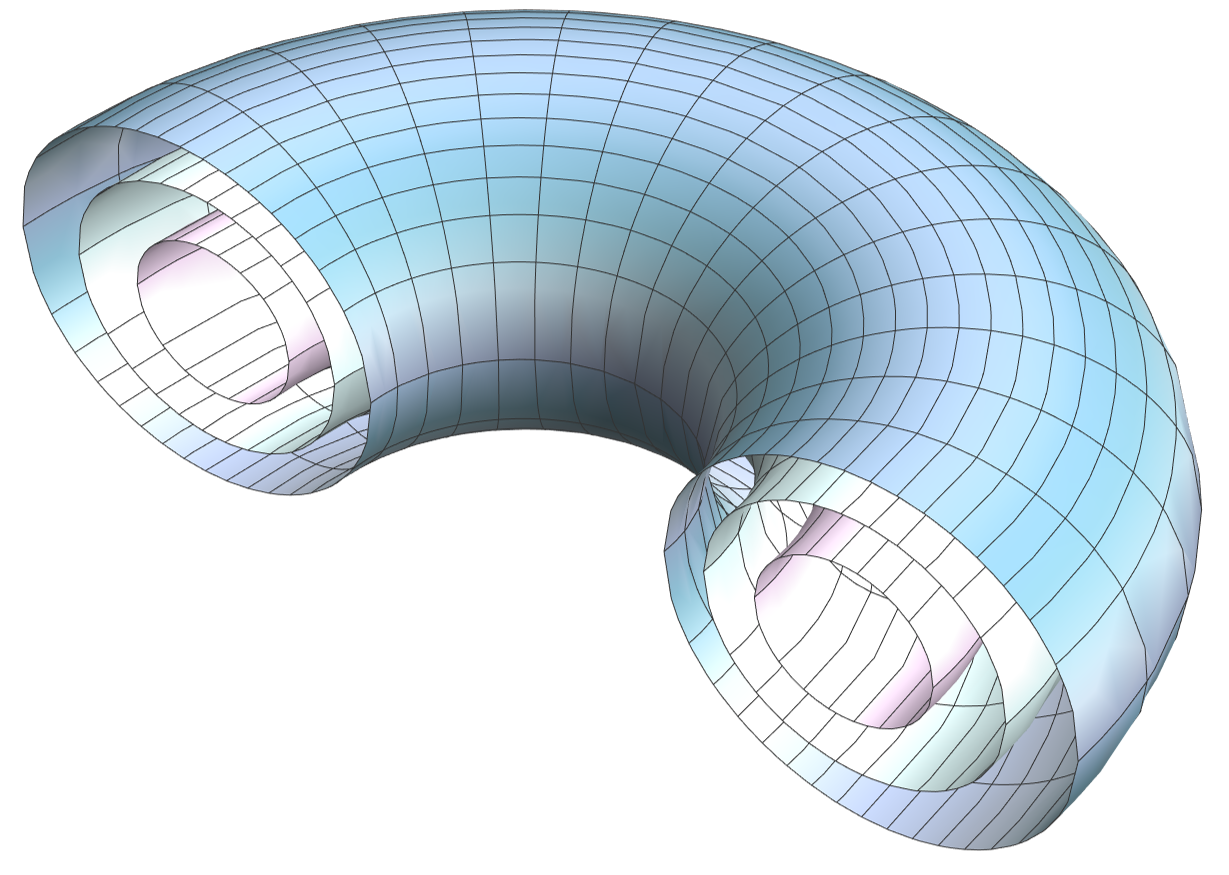}
   \label{fig:syn-Ergosphere02}}
   \quad
        \subfloat{
			\includegraphics[width=7cm]{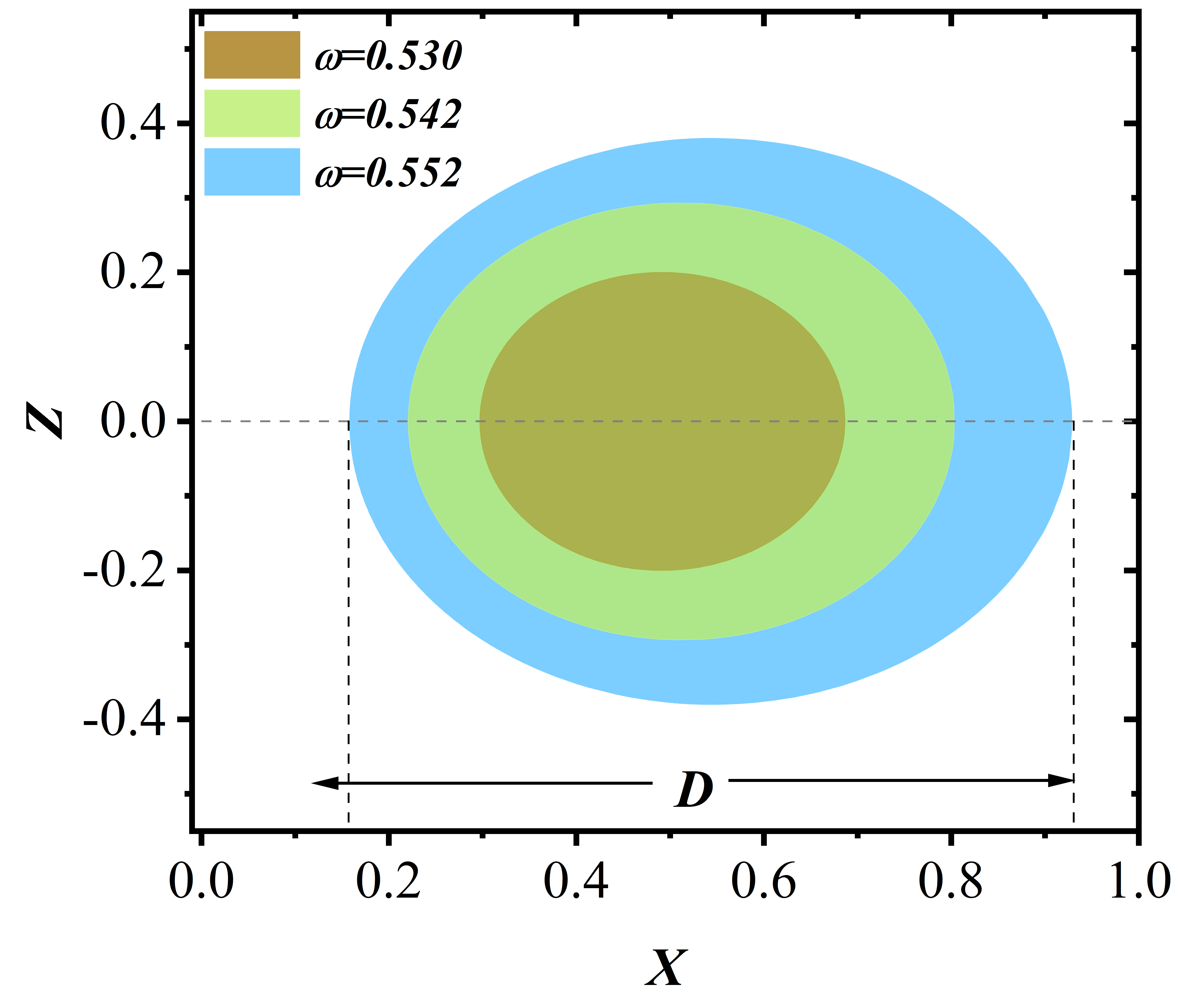}
   \label{fig:syn-Ergosphere01a}}
      \quad
		\subfloat{
			\includegraphics[width=7cm]{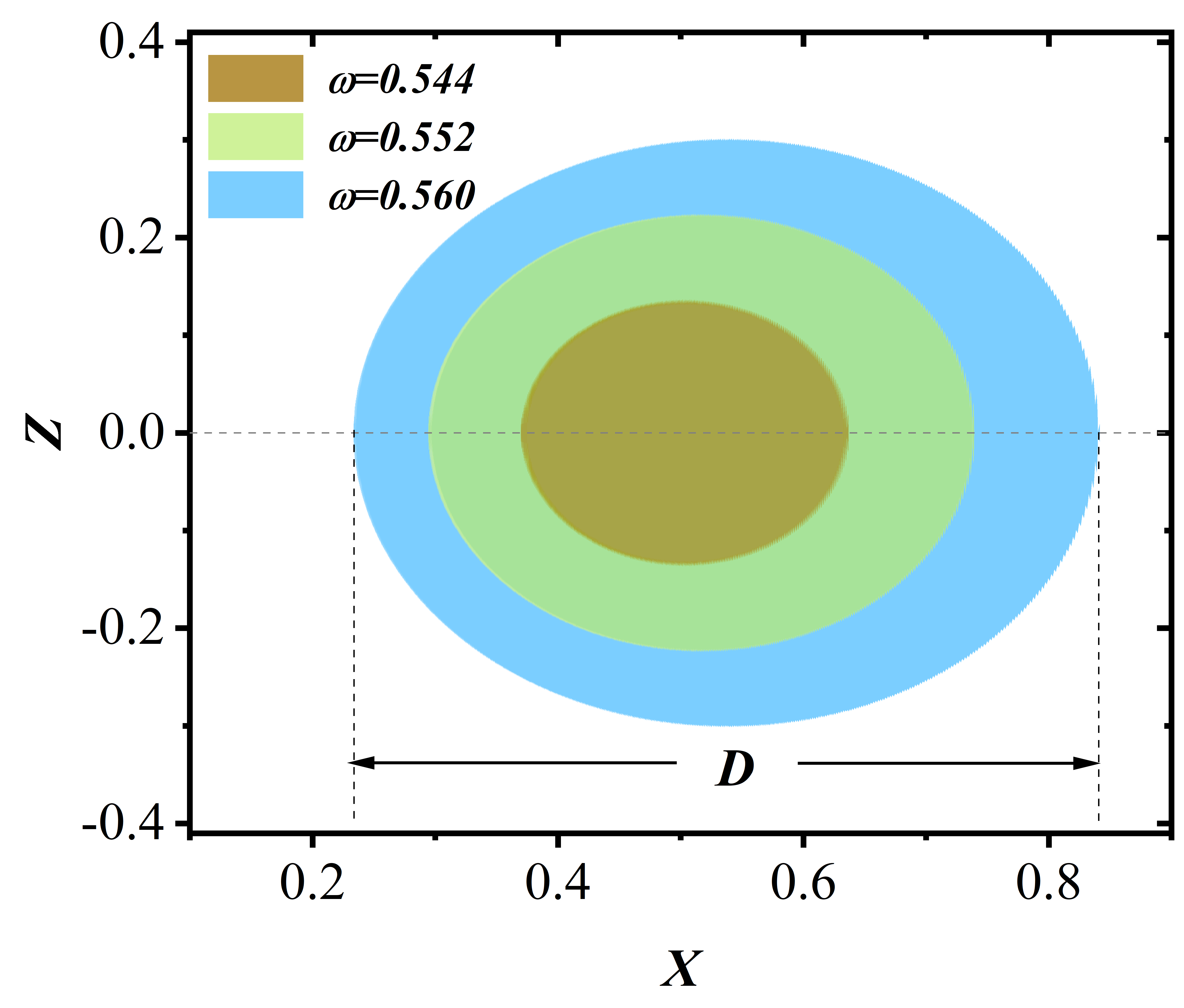}
   \label{fig:syn-Ergosphere02a}}
		\caption{Top left panel: For $\mu_{1}=0.69$, the ergosurface distribution of RMPSs with $\omega=0.530$ (purple), 0.542 (green), and 0.552 (blue). Top right panel: For $\mu_{1}=0.70$, the ergosurface distribution of RMPSs with $\omega=0.544$ (purple), 0.552 (green), and 0.560 (blue). Bottom left panel: For $\mu_{1}=0.69$, the cross-sectional views of ergosphere with respect to the XZ plane, with different colors corresponding to RMPSs with $\omega=0.530$ (brown), 0.542 (blue), and 0.552 (green). Bottom right panel: For $\mu_{1}=0.70$, the cross-sectional views of ergosphere with respect to the XZ plane, with different colors corresponding to RMPSs with $\omega=0.544$ (brown), 0.552 (blue), and 0.560 (green). }
		\label{fig:syn-Ergosphere}
	\end{figure}

    In Fig. \ref{fig:syn-E}, for a sufficiently large frequency $\mu_{1}$, the binding energy $E_m$ of RMPSs monotonically increases with increase of the synchronized frequency. However, when $\mu_{1}$ is sufficiently small, the binding energy of RMPSs initially increases and then decreases as the synchronized frequency increases, which is displayed in the embedded figure. When $\mu_{1}$ continues to decrease to a smaller value, the binding energy $E_m$ of RMPSs monotonically decreases with increase of the synchronized frequency. Moreover, analyzing the binding energy $E_{m}$, we find that the maximum value of the binding energy gradually decreases as $\mu_1$ decrease. When $\mu_{1}> 0.769$, the binding energy of RMPSs is always less than zero, the multistate solutions are always stable. When $\mu_{1}\leq 0.769$, the solutions can undergo a transition from a stable solution to an unstable one as the synchronized frequency $\omega$ increases under a fixed $\mu_{1}$. Tab. \ref{tab:syn-E} shows ranges of the synchronized frequency and binding energy of these solutions, as well as the values of the synchronized frequency corresponding to the maximum value, minimum value of $E_m$, and $E_m=0$, which are denoted by $\omega^{max}$, $\omega^{min}$, and $\omega^{0}$, respectively. 
    
    The study in Ref. \cite{Santos:2020pmh} depicts that there is ergosphere in rotating Proca stars, which is defined as the region where $g_{tt}$ is positive ($g_{tt}>0$). At the end of this subsection, we calculate the ergosphere of RMPSs. To facilitate illustration, we use Cartesian coordinates $X,Y,Z$ in our plots, which are related to the original coordinates by $X=(x\sin{\theta}\cos{\varphi})/(x-1),Y=(x\sin{\theta}\sin{\varphi})/(x-1),Z=(x\cos{\theta})/(x-1)$. 
    
    We first analyze the difference of ergosphere between the single-field Proca stars and RMPSs. The left panel of Fig. \ref{fig:syn-Ergosphere} depicts the ergosurface $(g_{tt}=0)$ of RMPSs (blue) under $\omega=0.56$, $\mu_{0}=1$, $\mu_{1}=0.7$. And the right panel shows the cross-sectional view of the ergosurface of RMPSs (green) in the $XZ$ plane. For convenience, we only display half of the ergosphere. To compare it with the single-field Proca stars, we also dipict the ground state Proca star under $\omega=0.56$, $\mu=1$. We also calculated the first excited state Proca star under $\omega=0.56$, $\mu=0.7$, however, it lacks ergosphere.

    From the Fig. \ref{fig:syn-Ergosphere}, it can be seen that the ergosphere $(g_{tt}>0)$ of RMPSs is similar to that of ground state Proca star, both are symmetric with respect to the equatorial plane, and the distribution range of ergosphere of RMPSs is smaller than that of ground state Proca star under the influence of the first excited state. In the right panel we exhibit the cross-sectional views of their ergosphere about the $XZ$ plane (right), where blue and green areas represent ergosphere of ground state Proca star and RMPSs, respectively.  
    \begin{figure}[htbp]
        \centering
        \includegraphics[width=7.5cm]{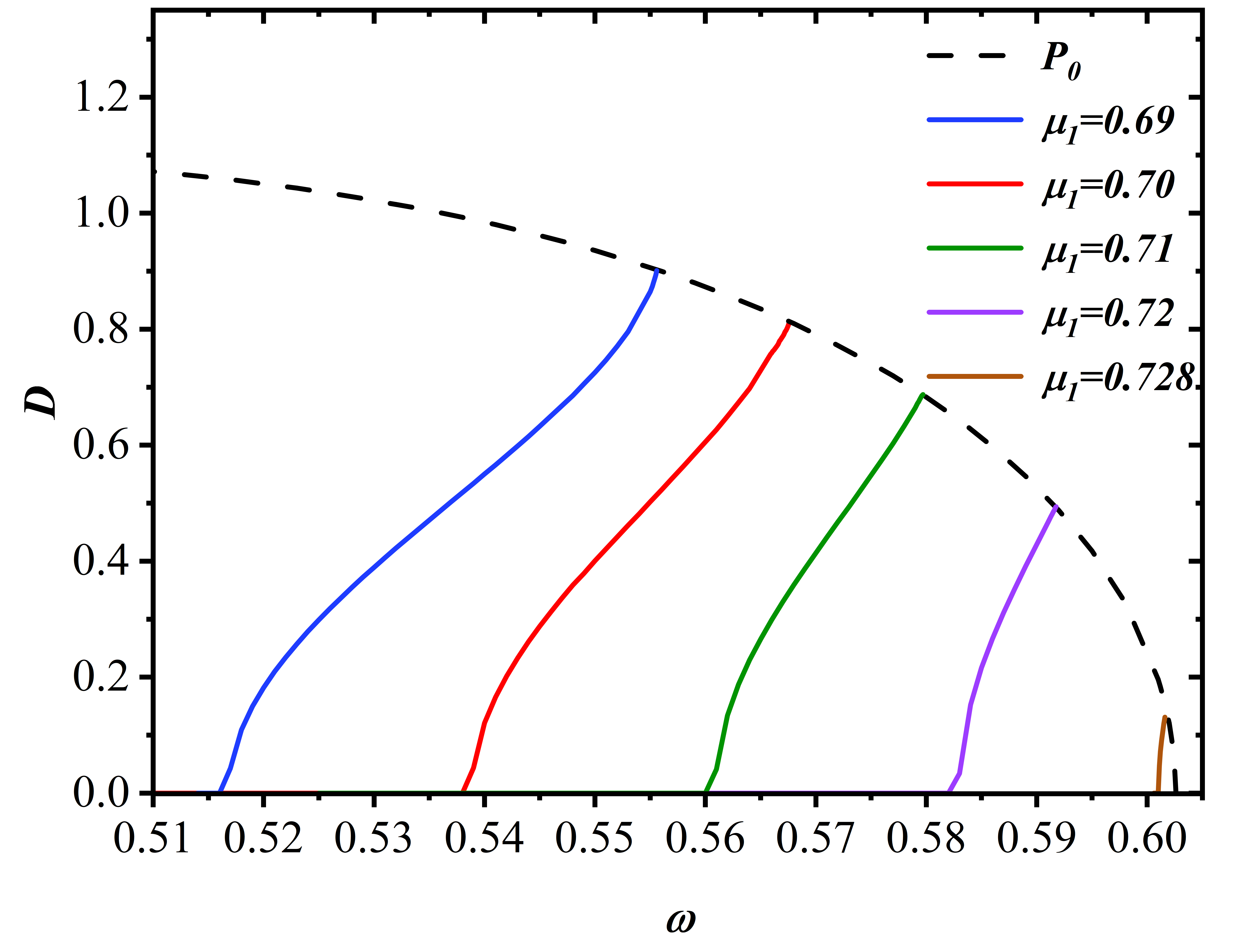}
        \caption{The value of $D$ as a fuction of the synchronized frequency $\omega$ for several $\mu_{1}$. The black dashed line represents the ground state Proca star, and the different colorful solid lines represent RMPSs with $\mu_{1}=0.69$ (blue), 0.70 (red), 0.71 (green), 0.72 (purple) and 0.728 (brown), respectively.}
        \label{fig:syn-D}
    \end{figure}

    Furthermore, we discuss how the ergosphere of RMPSs changes with synchronized frequency under a fixed value of $\mu_{1}$. Fig. \ref{fig:syn-Ergosphere} depicts the three-dimensional plots (top) of the ergosurface of RMPSs and its cross-sectional plots (bottom) of the ergospheres with respect to the $XZ$ plane for $\mu_{1}=0.69$ (left), $0.70$ (right), respectively. It can be seen that for a fixed $\mu_{1}$, the range of ergosphere is to increase with $\omega$. 
    
    To better described the change of the ergosphere range with synchronized frequency, we introduce the width of the ergosphere $D$ in the bottom panels of Fig. \ref{fig:syn-Ergosphere}, which is defined as the difference between the maximum and minimum distances of the ergosphere from the $z-axis$. Fig. \ref{fig:syn-D} depicts the value of $D$ as a function of synchronized frequency $\omega$ for $\mu_{1}=0.69$, 0.70, 0.71, 0.72, 0.728. We find that under a fixed $\mu_1$, the width of the ergospheres decreases as the frequency decreases and the width becomes zero for the sufficiently small frequency. Moreover, we can also analyze the influence of $\mu_1$ to the width of the ergosphere, obviously as $\mu_{1}$ increases, the maximum value of $D$ are to decrease, when $\mu_{1}>0.728$, there is no ergosphere in RMPSs.
    \begin{figure}[h!]
        \centering
        \subfloat{
        \includegraphics[width=7cm]{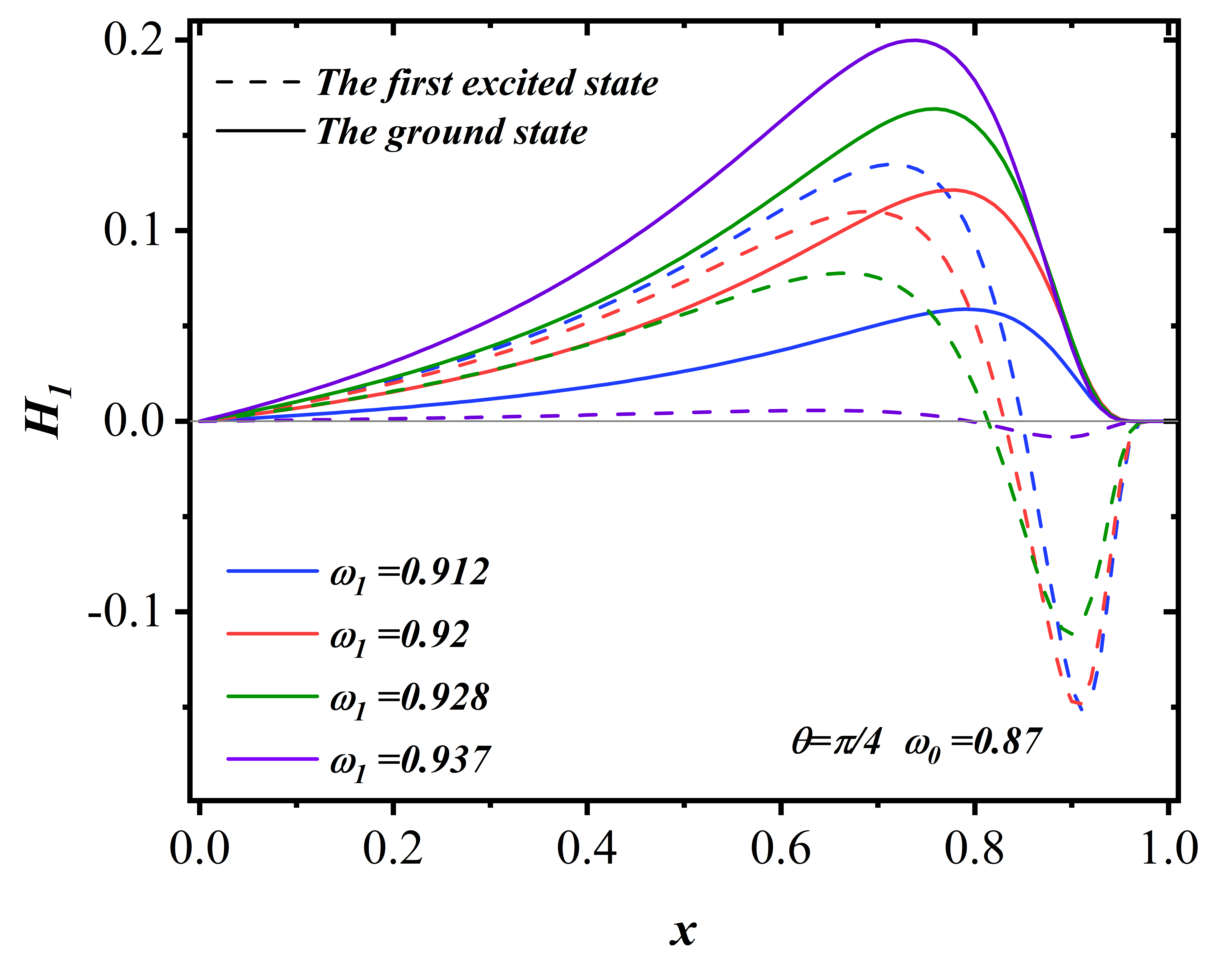}
        }
        \quad
        \subfloat{
        \includegraphics[width=7cm]{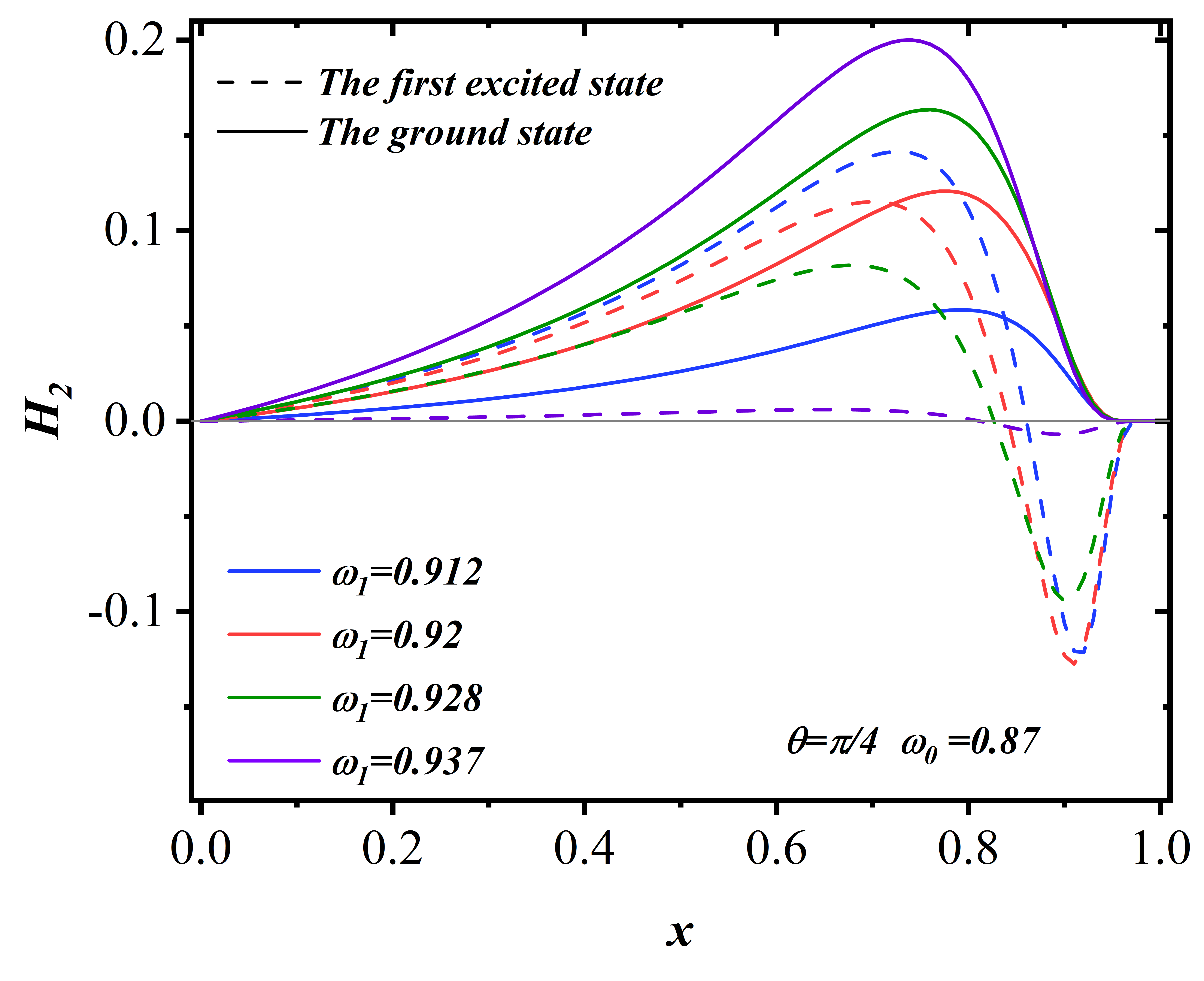}
        }
        \quad
        \subfloat{
        \includegraphics[width=7cm]{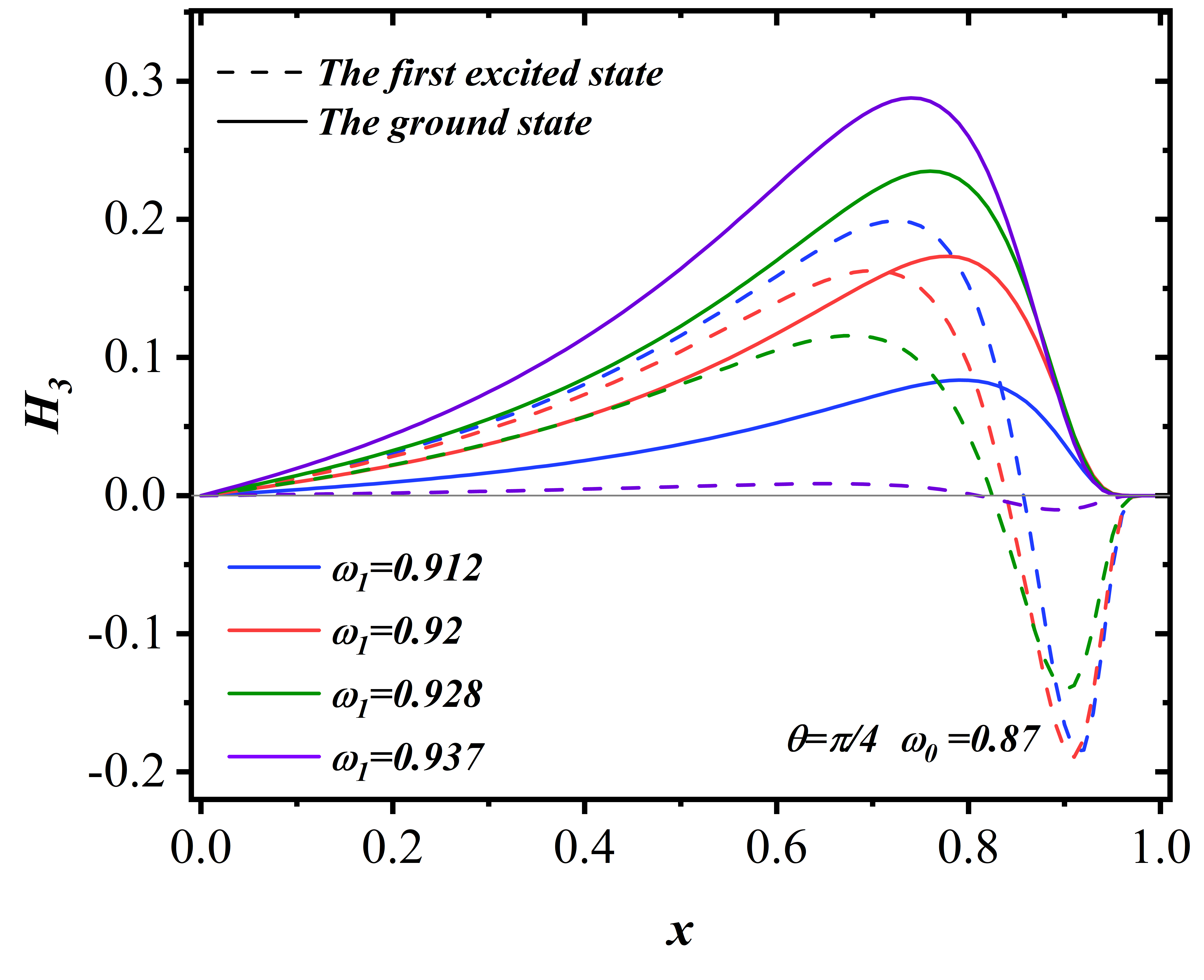}
        }
        \quad
        \subfloat{
        \includegraphics[width=7cm]{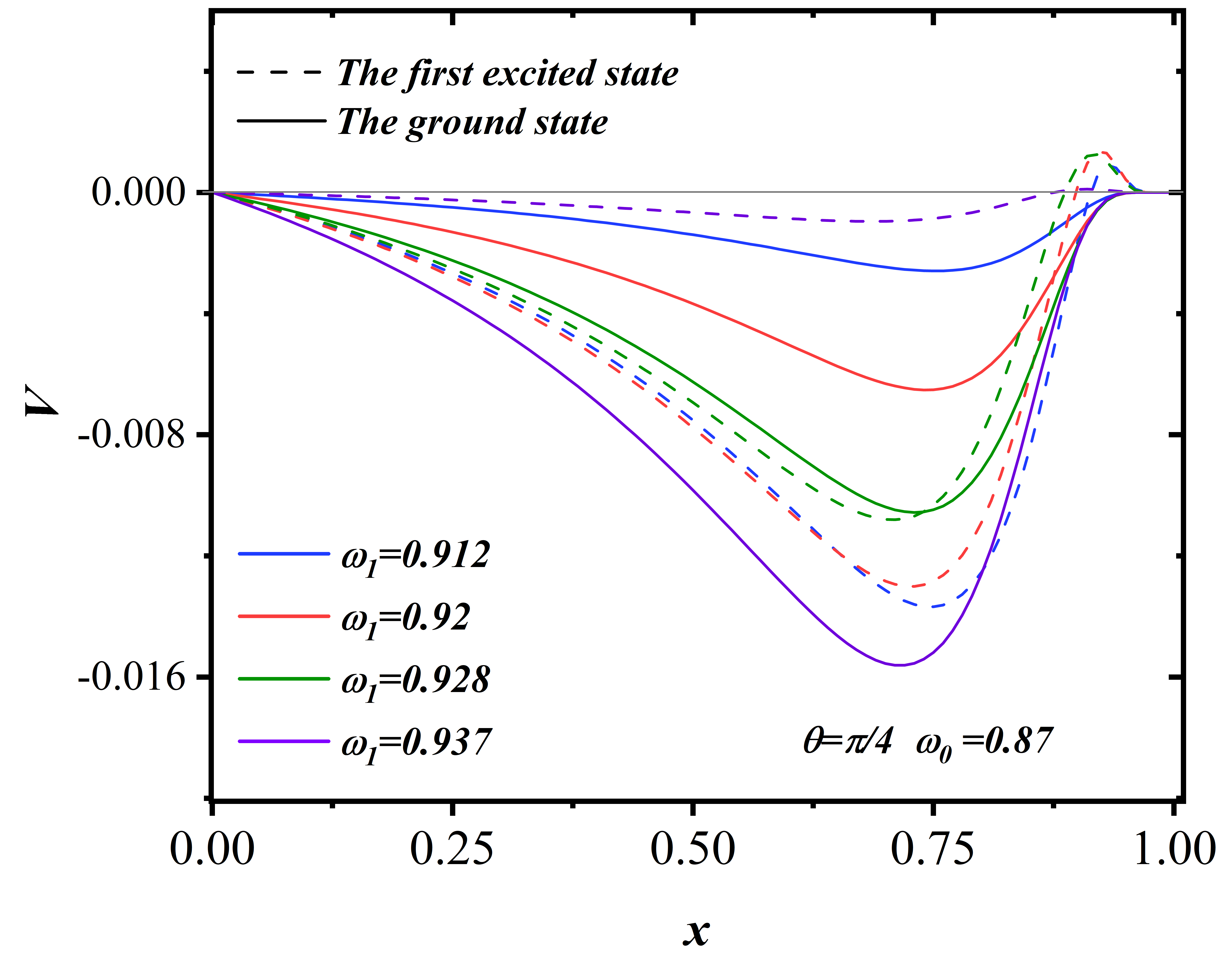}
        }
        \caption{The Proca fields distribution with respect to $x$ for severals different values of the non-synchronized frequency $\omega_1$ at $\theta=\frac{\pi}{4}$ when $\omega_{0}=0.87$. The solid line and the dashed line represent the ground state and the first excited state, respectively. Different colors represent $\omega_{1}=0.912$ (blue), $\omega_{1}=0.92$ (red), $\omega_{1}=0.928$ (green), and $\omega_{1}=0.937$ (purple). }
        \label{fig:non-syn-field}
    \end{figure}
    %

    \subsection{Non-synchronized frequency}
    In this subsection, we explore the case of RMPSs with non-synchronized frequencies. Unlike the synchronized frequency case, and the masses of the ground state and excited state can be equal. In order to better analyze the influence of other parameters on the solution family of non-synchronized frequencies, we set the mass of the excited state to 1 $(\mu_{0}=\mu_{1}=1)$, and investigate how the RMPSs change with respect to the excited state frequency $\omega_1$ with the ground state frequency $\omega_0$ fixed.

    Fig. \ref{fig:non-syn-field} depicts Proca fields distribution with respect to $x$ at $\theta=\pi/4$ under several values of non-synchronized frequency $\omega_{1}$, where the frequency $\omega_{0}$ of the ground state is set to 0.87. From the figure, it can be seen that similar to the case of synchronized frequency, $|{H}_j^{0}|_{max}$ and $|{V}^{0}|_{max}$ decrease as $\omega_{1}$ increases, while $|{H}_j^{1}|_{max}$ and $|{V}^{1}|_{max}$ both increase. As $\omega_{1}$ approaches its maximum value, $|{{H}_j^{0}}|_{max}$ and $|{V}^{0}|_{max}$ vanish, RMPSs degenerate into the $P_0$. Conversely, when the non-synchronized frequency decreases to its minimum value, RMPSs degenerate into the first excited state Proca star $P_1$.

    \begin{figure}[htbp]
        \subfloat{
        \centering
        \includegraphics[width=7cm]{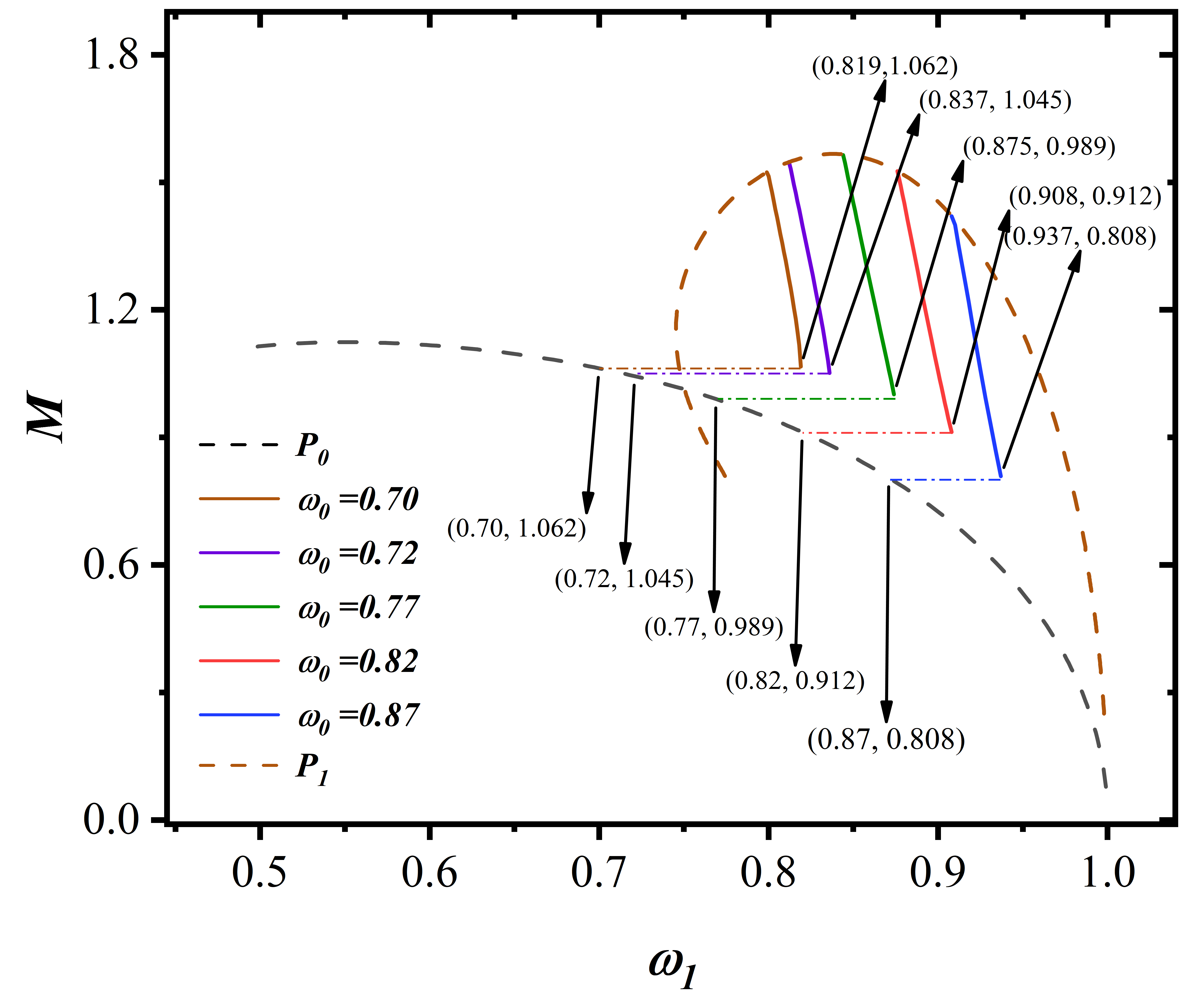}
        }
        \quad
        \subfloat{
        \centering
        \includegraphics[width=7cm]{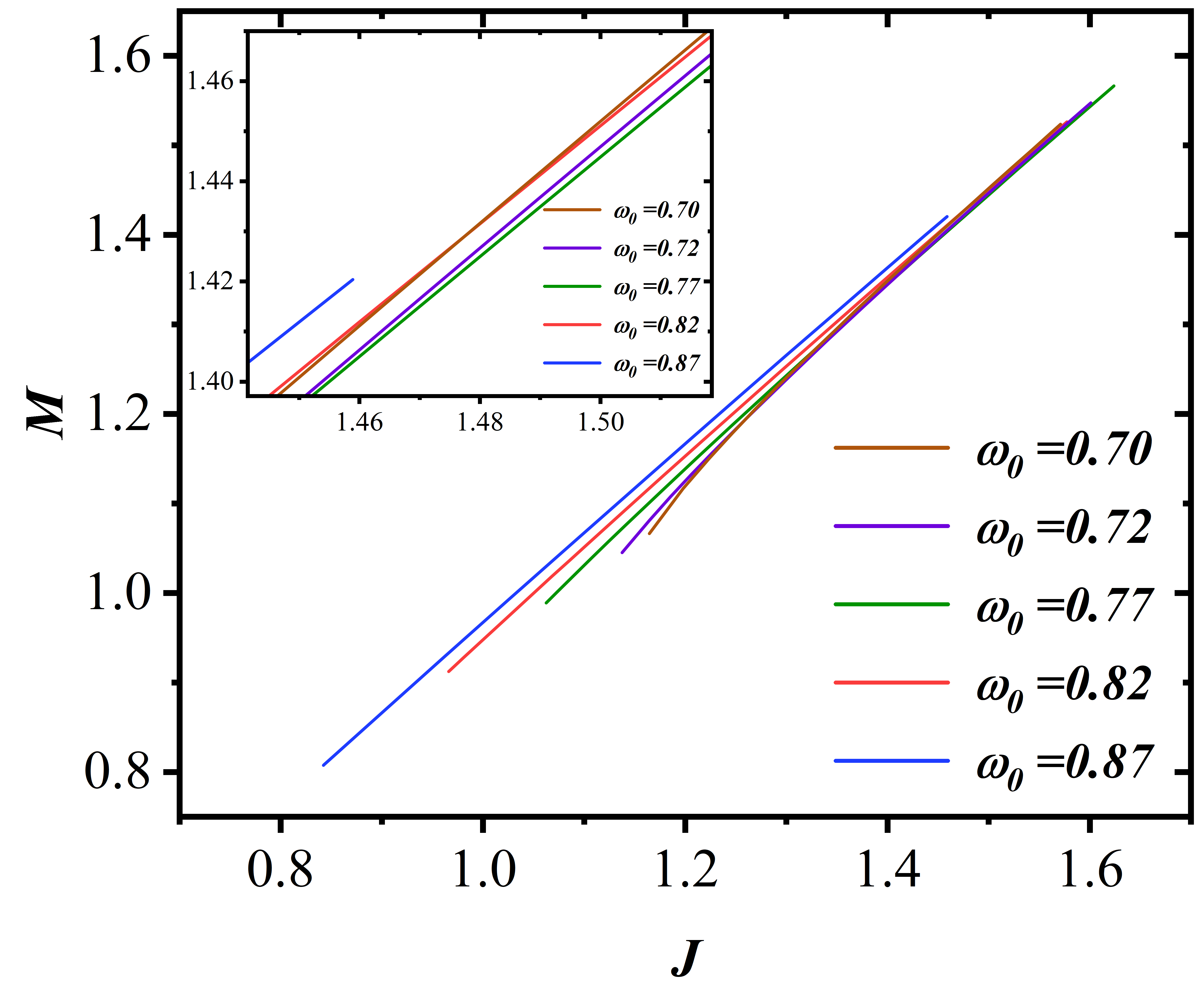}
        }
        \caption{Left panel: The ADM mass of non-synchronized frequency RMPSs (solid line) as a function of non-synchronized frequency $\omega_{1}$ for several different values of $\omega_{0}$. The different colorful solid line represent $\omega_{0}=0.70$ (brown), 0.72 (purple), 0.77 (green), 0.82 (red), 0.87 (blue), respectively. While the black dashed line represents the ground state Proca stars, and the brown dashed line represents the first excited state Proca stars. Right panel:  The ADM mass $M$ of synchronized RMPSs as a function of angular momentum $J$ for several different values of $\omega_{0}$. }
        \label{fig:nonsyn-J-M}
    \end{figure}

	To further discuss the properties of non-synchronized frequency RMPSs, we also calculated their ADM mass $M$ and angular momentum $J$. Fig. \ref{fig:nonsyn-J-M} depicts the ADM mass $M$ of non-synchronized RMPSs vs. $\omega_{1}$ (left) and the ADM mass $M$ vs. angular momentum $J$ (right) under five different $\omega_{0}$. The different colorful solid line represent $\omega_{0}=0.70$ (brown), 0.72 (purple), 0.77 (green), 0.82 (red), 0.87 (blue), respectively.  While the two dashed lines represent the ADM mass $M$ of the ground state and the excited state Proca stars as a function of their matter field frequencies. 
    \begin{table} [!b]
	\centering 
	\begin{tabular}{|c||c|c|c|}
		\hline
		&$\omega_1$ &$M$ &$J$     \\
		\hline
		$\omega_{0}=0.70$ & $0.799\sim0.819$ & $1.064\sim1.524$ & $1.164\sim1.571$\\ \hline
		$\omega_{0}=0.72$ & $0.812\sim0.836$ & $1.045\sim1.548$ & $1.138\sim1.601$\\ \hline
		$\omega_{0}=0.77$ & $0.843\sim0.875$ & $0.989\sim1.566$ & $1.062\sim1.634$\\ \hline
        $\omega_{0}=0.82$ & $0.876\sim0.908$ & $0.912\sim1.526$ & $0.966\sim1.577$\\ \hline   
        $\omega_{0}=0.87$ & $0.908\sim0.937$ & $0.808\sim1.420$ & $0.842\sim1.459$\\
		\hline
	\end{tabular}
 	\caption{The range of non-synchronized frequency $\omega_{1}$, ADM mass $M$, and angular momentum $J$. }
	\label{tab:non-syn-M-J}
    \end{table}

	From the left panel of Fig. \ref{fig:nonsyn-J-M}, it can be seen that under a fixed $\omega_{0}$, as $\omega_1$ increases, the ADM mass $M$ is to decrease monotonically, and finally interrupt at (0.819, 1.064), (0.836, 1.045), (0.875, 0.989), (0.908, 0.912), (0.937, 0.808), respectively, where the ADM masses are the same as those of the ground state Proca stars with $\omega=0.70$, $0.72$, $0.77$, $0.83$, and $0.87$. When $\omega_{1}$ tends to its minimum value, the curves intersect with the first excited state Proca star. In addition, according to the right panel of Fig. \ref{fig:nonsyn-J-M}, ADM mass $M$ of RMPSs increases monotonically with angular momentum $J$, and which is similar to the case with synchronized frequency. 
	
	To better analyze how the ADM mass and angular momentum of RMPSs change with $\omega_{0}$, in Tab. \ref{tab:non-syn-M-J}, we also show the range of non-synchronized frequency, ADM mass, and angular momentum. It can be seen that when $\omega_{0}$ increases, the minimum values of ADM mass and angular momentum decrease, while the maximum and minimum values of non-synchronized frequency $\omega_{1}$ increase.

    \begin{figure}[htbp]
        \centering
        \includegraphics[scale=0.35]{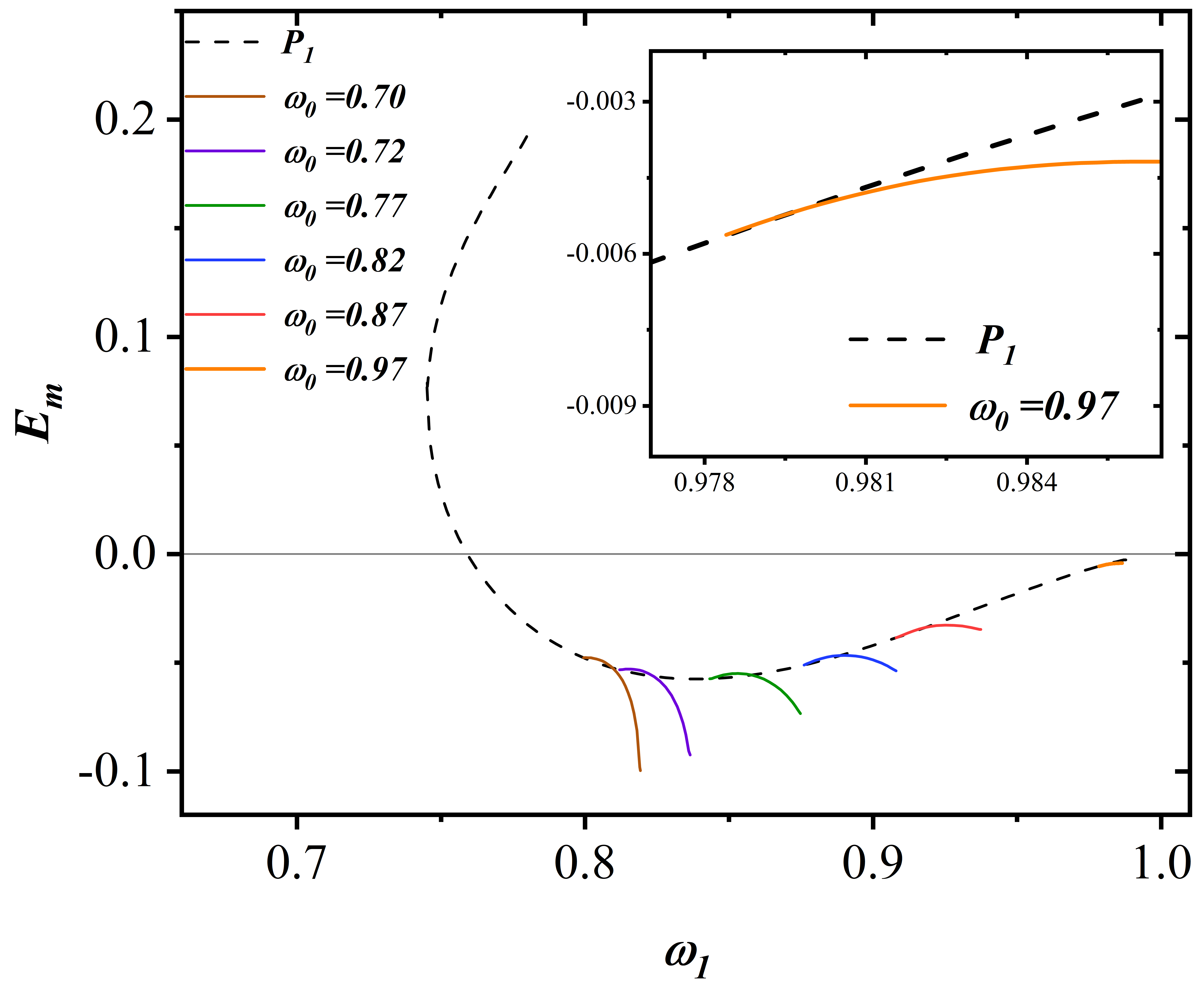}
        \caption{The binding energy of RMPSs $E_m$ as a fuction of $\omega_{1}$. Different colorful solid lines represent RMPSs with $\omega_{0}=0.7$ (brown), 0.72 (purple), 0.77 (green), 0.82 (blue), 0.87 (red) and 0.97 (orange), respectively. The black dashed line represents the binding energy of the first excited state Proca star $E_1$ as a function of its frequency. }
        \label{fig:non-syn-E}
    \end{figure}

    Furthermore, we also calculated the binding energy of the non-synchronized frequency RMPSs. Fig. \ref{fig:non-syn-E} depicts the distribution of the binding energy $E_m$ of the RMPSs with respect to $\omega_{1}$ when $\omega_{0}$ takes different values. The different colorful solid lines represent RMPSs with $\omega_{0}=0.7$ (brown), 0.72 (purple), 0.77 (green), 0.82 (blue), 0.87 (red) and 0.97 (orange), respectively. The binding energy of the first excited state Proca star as a function of its frequency is also shown in the figure, which is represented by the black dashed line. 
    
    As can be seen in Fig. \ref{fig:non-syn-E}, the binding energy $E_m$ of RMPSs monotonically increases with increase of the synchronized frequency for a sufficiently large $\omega_{0}$, similar to the case with synchronized frequency. When $\omega_{0}$ decreases to a sufficiently small value, the binding energy $E_m$ of multifield solutions first increases and then decreases as $\omega_{1}$ increases. As $\omega_{0}$ is reduced to a smaller value, the binding energy $E_m$ monotonically decreases with the increase of the synchronized frequency, as $\omega_{0}=0.70$ shown in this image. When $\omega_{0}\geq 0.70$, all non-synchronized frequency solutions we obtained are stable. However, when $\omega_{0}$ continues to decrease, it's very hard to find solutions. Tab. \ref{tab:non-syn-E} depicts in detail the range of $\omega_{1}$ and $E_m$, as well as the values of non-synchronized frequency $\omega_{1}^{max}$ and $\omega_{1}^{min}$ corresponding to the maximum and minimum values of $E_m$, respectively. Finally, we also calculated the $g_{tt}$ of the non-synchronized frequency RMPSs, there is no ergosphere in these multistate solutions. 
    \begin{table}[htbp]
	\centering 
	\begin{tabular}{|c||c|c|c|c|}
		\hline
		&$\omega_{1}$ &$E$ & $\omega_{1}^{min}$ & $\omega_{1}^{max}$\\
		\hline
		$\omega_{0}=0.70$ & $0.799\sim0.819$ & $-1.000\sim-0.047$ & 0.819 & 0.799\\ \hline
		$\omega_{0}=0.72$ & $0.812\sim0.837$ & $-0.092\sim-0.053$ & 0.837 & 0.815\\ \hline
		$\omega_{0}=0.77$ & $0.843\sim0.875$ & $-0.073\sim-0.055$ & 0.875 & 0.853\\ \hline
        $\omega_{0}=0.82$ & $0.876\sim0.908$ & $-0.054\sim-0.047$ & 0.908 & 0.890\\ \hline
        $\omega_{0}=0.87$ & $0.908\sim0.937$ & $-0.035\sim-0.033$ & 0.908 & 0.925\\ \hline
        $\omega_{0}=0.97$ & $0.978\sim0.987$ & $-0.006\sim-0.004$ & 0.978 & 0.987\\
		\hline
	\end{tabular}
    \caption{The range of the non-synchronized frequency $\omega_{1}$ and binding energy $E_m$, as well as the value of the non-synchronized frequency $\omega_{1}$ corresponding to the maximum value, minimum value of $E_m$. }
    \label{tab:non-syn-E}
\end{table}
%
	%

	\section{CONCLUSION}
	\label{sec5}		
    In this study, we constructed rotating multistate Proca stars (RMPSs) the coexisting states of two Proca fields, includingthat the ground state and the first excited state. We calculated RMPSs in synchronized frequency and non-synchronized frequency scenarios and subsequently analyzed the distribution of their matter field function.  We observed that the maximum absolute values of the first excited state monotonically decreases as $\omega(\omega_1)$ increases, and at the minimum and maximum frequencies $\omega(\omega_1)$, the RMPSs degenerates into the ground state Proca star or the first excited state Proca star.  \par
    To further investigate the properties of multistate Proca stars, we calculated their ADM mass $M$, angular momentum $J$, and binding energy $E_{m}$. The calculation results show that for the synchronized frequency RMPSs, as $\mu_{1}$ decreases, the minimum ADM mass of RMPSs increase, and the maximum and minimum synchronized frequencies $\omega$ decrease. In the case of non-synchronized frequency RMPSs, as $\omega_{0}$ decreases, the minimum values of ADM mass $M$ and angular momentum increase, and the maximum and minimum values of non-synchronized frequency $\omega_{1}$ decrease. For both two types of RMPSs, the ADM mass M monotonically increases with increasing $J$. Moreover, analyzing the binding energy $E_{m}$, we find that for synchronized frequency RMPSs, the solutions are stable under a sufficiently large $\mu_{1}$, while the unstable solution appears for $\mu_{1}\leq 0.769$.\par
    Furthermore, we investigated the ergosphere of RMPSs and compared our findings with those of single-field Proca stars. We found that the ergosphere of synchronized frequency RMPSs appears when $\mu_{1}\leq 0.728$, the width of the ergospheres also increase as synchronized frequency $\omega$ increases for a fixed $\mu_{1}$. And as $\mu_{1}$ decreases, the maximum value of the width of the ergospheres increases. Moreover, due to the influence of the first excited state, the width of the ergosphere of multifield solutions is smaller than that of ground state Proca stars. \par
    In this study, the azimuthal harmonic indexes are set to 1 $(m_0=m_1=1)$. Our future plans involve extending our work to RMPSs with a greater number of azimuthal harmonic index. Furthermore, the multistate solution construction method we used in this study can be applied to other matter fields. In our upcoming research, we aim to generalize our research to multifield models coexisting Proca fields and different matter fields.\par

	\section*{ACKNOWLEDGEMENTS}
	This work is supported by National Key Research and Development Program of China (Grant No. 2020YFC2201503) and the National Natural Science Foundation of China (Grants No.~12275110 and No.~12247101).

		\begin{small}

	\end{small}
	

\begin{thebibliography}{99}
\bibitem{Sharma:2008sc}
R.~Sharma, S.~Karmakar and S.~Mukherjee,
``Boson star and dark matter,''
[arXiv:0812.3470 [gr-qc]].

\bibitem{Eby:2015hsq}
J.~Eby, C.~Kouvaris, N.~G.~Nielsen and L.~C.~R.~Wijewardhana,
``Boson Stars from Self-Interacting Dark Matter,''
JHEP \textbf{02}, 028 (2016)
[arXiv:1511.04474 [hep-ph]].

\bibitem{Chen:2020cef}
J.~Chen, X.~Du, E.~W.~Lentz, D.~J.~E.~Marsh and J.~C.~Niemeyer,
``New insights into the formation and growth of boson stars in dark matter halos,''
Phys. Rev. D \textbf{104}, no.8, 083022 (2021)
[arXiv:2011.01333 [astro-ph.CO]].

\bibitem{Gorghetto:2022sue}
M.~Gorghetto, E.~Hardy, J.~March-Russell, N.~Song and S.~M.~West,
``Dark photon stars: formation and role as dark matter substructure,''
JCAP \textbf{08}, no.08, 018 (2022)
[arXiv:2203.10100 [hep-ph]].

\bibitem{Vincent:2015xta}
F.~H.~Vincent, Z.~Meliani, P.~Grandclement, E.~Gourgoulhon and O.~Straub,
``Imaging a boson star at the Galactic center,''
Class. Quant. Grav. \textbf{33}, no.10, 105015 (2016)
[arXiv:1510.04170 [gr-qc]].

\bibitem{Cunha:2016bjh}
P.~V.~P.~Cunha, J.~Grover, C.~Herdeiro, E.~Radu, H.~Runarsson and A.~Wittig,
``Chaotic lensing around boson stars and Kerr black holes with scalar hair,''
Phys. Rev. D \textbf{94}, no.10, 104023 (2016)
[arXiv:1609.01340 [gr-qc]].

\bibitem{Wheeler:1955zz}
J.~A.~Wheeler,
``Geons,''
Phys. Rev. \textbf{97} (1955), 511-536

\bibitem{Power:1957zz}
E.~A.~Power and J.~A.~Wheeler,
``Thermal Geons,''
Rev. Mod. Phys. \textbf{29} (1957), 480-495

\bibitem{Palenzuela:2017kcg}
C.~Palenzuela, P.~Pani, M.~Bezares, V.~Cardoso, L.~Lehner and S.~Liebling,
``Gravitational Wave Signatures of Highly Compact Boson Star Binaries,''
Phys. Rev. D \textbf{96}, no.10, 104058 (2017)
[arXiv:1710.09432 [gr-qc]].

\bibitem{Bezares:2017mzk}
M.~Bezares, C.~Palenzuela and C.~Bona,
``Final fate of compact boson star mergers,''
Phys. Rev. D \textbf{95}, no.12, 124005 (2017)
[arXiv:1705.01071 [gr-qc]].

\bibitem{Bezares:2018qwa}
M.~Bezares and C.~Palenzuela,
``Gravitational Waves from Dark Boson Star binary mergers,''
Class. Quant. Grav. \textbf{35}, no.23, 234002 (2018)
[arXiv:1808.10732 [gr-qc]].

\bibitem{Jaramillo:2022zwg}
V.~Jaramillo, N.~Sanchis-Gual, J.~Barranco, A.~Bernal, J.~C.~Degollado, C.~Herdeiro, M.~Megevand and D.~N\'u\~nez,
``Head-on collisions of \ensuremath{\ell}-boson stars,''
Phys. Rev. D \textbf{105}, no.10, 104057 (2022)
[arXiv:2202.00696 [gr-qc]].

\bibitem{Kaup:1968zz}
D.~J.~Kaup,
``Klein-Gordon Geon,''
Phys. Rev. \textbf{172}, 1331-1342 (1968)

\bibitem{Ruffini:1969qy}
R.~Ruffini and S.~Bonazzola,
``Systems of selfgravitating particles in general relativity and the concept of an equation of state,''
Phys. Rev. \textbf{187} (1969), 1767-1783

\bibitem{Schunck:1996he}
F.~E.~Schunck and E.~W.~Mielke,
``Rotating boson star as an effective mass torus in general relativity,''
Phys. Lett. A \textbf{249} (1998), 389-394

\bibitem{Yoshida:1997qf}
S.~Yoshida and Y.~Eriguchi,
``Rotating boson stars in general relativity,''
Phys. Rev. D \textbf{56} (1997), 762-771

\bibitem{Kleihaus:2005me}
B.~Kleihaus, J.~Kunz and M.~List,
``Rotating boson stars and Q-balls,''
Phys. Rev. D \textbf{72}, 064002 (2005)
[arXiv:gr-qc/0505143 [gr-qc]].

\bibitem{Li:2019mlk}
H.~B.~Li, S.~Sun, T.~T.~Hu, Y.~Song and Y.~Q.~Wang,
``Rotating multistate boson stars,''
Phys. Rev. D \textbf{101}, no.4, 044017 (2020)
[arXiv:1906.00420 [gr-qc]].

\bibitem{Siemonsen:2020hcg}
N.~Siemonsen and W.~E.~East,
``Stability of rotating scalar boson stars with nonlinear interactions,''
Phys. Rev. D \textbf{103}, no.4, 044022 (2021)
[arXiv:2011.08247 [gr-qc]].

\bibitem{Astefanesei:2003qy}
D.~Astefanesei and E.~Radu,
``Boson stars with negative cosmological constant,''
Nucl. Phys. B \textbf{665} (2003), 594-622
[arXiv:gr-qc/0309131 [gr-qc]].

\bibitem{Prikas:2004yw}
A.~Prikas,
``Q stars in anti-de Sitter space-time,''
Gen. Rel. Grav. \textbf{36} (2004), 1841-1869
[arXiv:hep-th/0403019 [hep-th]].

\bibitem{Buchel:2013uba}
A.~Buchel, S.~L.~Liebling and L.~Lehner,
``Boson stars in AdS spacetime,''
Phys. Rev. D \textbf{87} (2013) no.12, 123006
[arXiv:1304.4166 [gr-qc]].

\bibitem{Jetzer:1989av}
P.~Jetzer and J.~J.~van der Bij,
``CHARGED BOSON STARS,''
Phys. Lett. B \textbf{227} (1989), 341-346


\bibitem{Brihaye:2014gua}
Y.~Brihaye, V.~Diemer and B.~Hartmann,
``Charged Q-balls and boson stars and dynamics of charged test particles,''
Phys. Rev. D \textbf{89} (2014) no.8, 084048
[arXiv:1402.1055 [gr-qc]].

\bibitem{Jetzer:1992tog}
P.~Jetzer, P.~Liljenberg and B.~S.~Skagerstam,
``Charged boson stars and vacuum instabilities,''
Astropart. Phys. \textbf{1} (1993), 429-448
[arXiv:astro-ph/9305014 [astro-ph]].

\bibitem{Kling:2017hjm}
F.~Kling and A.~Rajaraman,
``Profiles of boson stars with self-interactions,''
Phys. Rev. D \textbf{97}, no.6, 063012 (2018)
[arXiv:1712.06539 [hep-ph]].

\bibitem{Schunck:1999zu}
F.~E.~Schunck and D.~F.~Torres,
``Boson stars with generic selfinteractions,''
Int. J. Mod. Phys. D \textbf{9}, 601-618 (2000)
[arXiv:gr-qc/9911038 [gr-qc]].

\bibitem{Sanchis-Gual:2021phr}
N.~Sanchis-Gual, C.~Herdeiro and E.~Radu,
``Self-interactions can stabilize excited boson stars,''
Class. Quant. Grav. \textbf{39}, no.6, 064001 (2022)
[arXiv:2110.03000 [gr-qc]].

\bibitem{Herdeiro:2014goa}
C.~A.~R.~Herdeiro and E.~Radu,
``Kerr black holes with scalar hair,''
Phys. Rev. Lett. \textbf{112} (2014), 221101
[arXiv:1403.2757 [gr-qc]].

\bibitem{Herdeiro:2015waa}
C.~A.~R.~Herdeiro and E.~Radu,
``Asymptotically flat black holes with scalar hair: a review,''
Int. J. Mod. Phys. D \textbf{24} (2015) no.09, 1542014
[arXiv:1504.08209 [gr-qc]].

\bibitem{Herdeiro:2015gia}
C.~Herdeiro and E.~Radu,
``Construction and physical properties of Kerr black holes with scalar hair,''
Class. Quant. Grav. \textbf{32} (2015) no.14, 144001
[arXiv:1501.04319 [gr-qc]].

\bibitem{Herdeiro:2015tia}
C.~A.~R.~Herdeiro, E.~Radu and H.~R\'unarsson,
``Kerr black holes with self-interacting scalar hair: hairier but not heavier,''
Phys. Rev. D \textbf{92} (2015) no.8, 084059
[arXiv:1509.02923 [gr-qc]].

\bibitem{Cunha:2016bpi}
P.~V.~P.~Cunha, C.~A.~R.~Herdeiro, E.~Radu and H.~F.~Runarsson,
Int. J. Mod. Phys. D \textbf{25} (2016) no.09, 1641021
[arXiv:1605.08293 [gr-qc]].

\bibitem{Delgado:2020hwr}
J.~F.~M.~Delgado, C.~A.~R.~Herdeiro and E.~Radu,
``Kerr black holes with synchronized axionic hair,''
Phys. Rev. D \textbf{103} (2021) no.10, 104029
[arXiv:2012.03952 [gr-qc]].

\bibitem{Rosen:1994rq}
N.~Rosen,
``A Classical Proca particle,''
Found. Phys. \textbf{24} (1994), 1689-1695

\bibitem{Brito:2015pxa}
R.~Brito, V.~Cardoso, C.~A.~R.~Herdeiro and E.~Radu,
``Proca stars: Gravitating Bose\textendash{}Einstein condensates of massive spin 1 particles,''
Phys. Lett. B \textbf{752}, 291-295 (2016)
[arXiv:1508.05395 [gr-qc]].

\bibitem{Herdeiro:2017phl}
C.~A.~R.~Herdeiro and E.~Radu,
``Dynamical Formation of Kerr Black Holes with Synchronized Hair: An Analytic Model,''
Phys. Rev. Lett. \textbf{119} (2017) no.26, 261101
[arXiv:1706.06597 [gr-qc]].

\bibitem{SalazarLandea:2016bys}
I.~Salazar Landea and F.~Garc\'\i{}a,
``Charged Proca Stars,''
Phys. Rev. D \textbf{94}, no.10, 104006 (2016)
[arXiv:1608.00011 [hep-th]].

\bibitem{Duarte:2016lig}
M.~Duarte and R.~Brito,
``Asymptotically anti-de Sitter Proca Stars,''
Phys. Rev. D \textbf{94} (2016) no.6, 064055
[arXiv:1609.01735 [gr-qc]].

\bibitem{Minamitsuji:2018kof}
M.~Minamitsuji,
``Vector boson star solutions with a quartic order self-interaction,''
Phys. Rev. D \textbf{97}, no.10, 104023 (2018)
[arXiv:1805.09867 [gr-qc]].

\bibitem{Herdeiro:2023lze}
C.~Herdeiro, E.~Radu and E.~dos Santos Costa Filho,
``Proca-Higgs balls and stars in a UV completion for Proca self-interactions,''
JCAP \textbf{05}, 022 (2023)
[arXiv:2301.04172 [gr-qc]].

\bibitem{Sanchis-Gual:2019ljs}
N.~Sanchis-Gual, F.~Di Giovanni, M.~Zilh\~ao, C.~Herdeiro, P.~Cerd\'a-Dur\'an, J.~A.~Font and E.~Radu,
``Nonlinear Dynamics of Spinning Bosonic Stars: Formation and Stability,''
Phys. Rev. Lett. \textbf{123}, no.22, 221101 (2019)
[arXiv:1907.12565 [gr-qc]].

\bibitem{DiGiovanni:2020ror}
F.~Di Giovanni, N.~Sanchis-Gual, P.~Cerd\'a-Dur\'an, M.~Zilh\~ao, C.~Herdeiro, J.~A.~Font and E.~Radu,
``Dynamical bar-mode instability in spinning bosonic stars,''
Phys. Rev. D \textbf{102}, no.12, 124009 (2020)
[arXiv:2010.05845 [gr-qc]].

\bibitem{Herdeiro:2021lwl}
C.~A.~R.~Herdeiro, A.~M.~Pombo, E.~Radu, P.~V.~P.~Cunha and N.~Sanchis-Gual,
``The imitation game: Proca stars that can mimic the Schwarzschild shadow,''
JCAP \textbf{04}, 051 (2021)
[arXiv:2102.01703 [gr-qc]].

\bibitem{Sanchis-Gual:2017bhw}
N.~Sanchis-Gual, C.~Herdeiro, E.~Radu, J.~C.~Degollado and J.~A.~Font,
``Numerical evolutions of spherical Proca stars,''
Phys. Rev. D \textbf{95} (2017) no.10, 104028
[arXiv:1702.04532 [gr-qc]].

\bibitem{CalderonBustillo:2020fyi}
J.~Calder\'on Bustillo, N.~Sanchis-Gual, A.~Torres-Forn\'e, J.~A.~Font, A.~Vajpeyi, R.~Smith, C.~Herdeiro, E.~Radu and S.~H.~W.~Leong,
``GW190521 as a Merger of Proca Stars: A Potential New Vector Boson of $8.7\times 10^{-13}$  eV,''
Phys. Rev. Lett. \textbf{126}, no.8, 081101 (2021)
[arXiv:2009.05376 [gr-qc]].

\bibitem{Ma:2023vfa}
T.~X.~Ma, C.~Liang, J.~Yang and Y.~Q.~Wang,
``Hybrid Proca-boson stars,''
[arXiv:2304.08019 [gr-qc]].

\bibitem{Pombo:2023sih}
A.~M.~Pombo, J.~M.~S.~Oliveira and N.~M.~Santos,
``Coupled scalar-Proca soliton stars,''
Phys. Rev. D \textbf{108} (2023) no.4, 044044

\bibitem{Bernal:2009zy}
A.~Bernal, J.~Barranco, D.~Alic and C.~Palenzuela,
``Multi-state Boson Stars,''
Phys. Rev. D \textbf{81}, 044031 (2010)
[arXiv:0908.2435 [gr-qc]].

\bibitem{Li:2020ffy}
H.~B.~Li, Y.~B.~Zeng, Y.~Song and Y.~Q.~Wang,
``Self-interacting multistate boson stars,''
JHEP \textbf{04}, 042 (2021)
[arXiv:2006.11281 [gr-qc]].

\bibitem{Guo:2020tla}
H.~K.~Guo, K.~Sinha, C.~Sun, J.~Swaim and D.~Vagie,
``Two-scalar Bose-Einstein condensates: from stars to galaxies,''
JCAP \textbf{10}, 028 (2021)
[arXiv:2010.15977 [astro-ph.CO]].

\bibitem{Sanchis-Gual:2021edp}
N.~Sanchis-Gual, F.~Di Giovanni, C.~Herdeiro, E.~Radu and J.~A.~Font,
``Multifield, Multifrequency Bosonic Stars and a Stabilization Mechanism,''
Phys. Rev. Lett. \textbf{126}, no.24, 241105 (2021)
[arXiv:2103.12136 [gr-qc]].

\bibitem{Sun:2022duv}
S.~X.~Sun, L.~Zhao and Y.~Q.~Wang,
``Chains of mini-boson stars,''
JHEP \textbf{08}, 152 (2023)
[arXiv:2210.09265 [gr-qc]].

\bibitem{Zeng:2023hvq}
Y.~B.~Zeng, S.~X.~Sun, S.~Y.~Cui, Y.~P.~Zhang and Y.~Q.~Wang,
``Rotating multistate axion boson stars,''
[arXiv:2309.05743 [gr-qc]].

\bibitem{Liang:2022mjo}
C.~Liang, J.~R.~Ren, S.~X.~Sun and Y.~Q.~Wang,
``Dirac-boson stars,''
JHEP \textbf{02} (2023), 249
[arXiv:2207.11147 [gr-qc]].

\bibitem{Huang:2023glq}
L.~X.~Huang, S.~X.~Sun, R.~Zhang, C.~Liang and Y.~Q.~Wang,
``Excited Dirac stars with higher azimuthal harmonic index,''
[arXiv:2309.16497 [gr-qc]].

\bibitem{Herdeiro:2016tmi}
C.~Herdeiro, E.~Radu and H.~R\'unarsson,
``Kerr black holes with Proca hair,''
Class. Quant. Grav. \textbf{33} (2016) no.15, 154001
[arXiv:1603.02687 [gr-qc]].

\bibitem{Herdeiro:2020jzx}
C.~A.~R.~Herdeiro and E.~Radu,
``Asymptotically flat, spherical, self-interacting scalar, Dirac and Proca stars,''
Symmetry \textbf{12} (2020) no.12, 2032
[arXiv:2012.03595 [gr-qc]].

\bibitem{Santos:2020pmh}
N.~M.~Santos, C.~L.~Benone, L.~C.~B.~Crispino, C.~A.~R.~Herdeiro and E.~Radu,
``Black holes with synchronised Proca hair: linear clouds and fundamental non-linear solutions,''
JHEP \textbf{07} (2020), 010
[arXiv:2004.09536 [gr-qc]].
			
		\end{thebibliography}
\end{document}